\documentclass[a4paper,fleqn,usenatbib]{mnras}

\usepackage{newtxtext,newtxmath}

\usepackage[T1]{fontenc}
\usepackage{ae,aecompl}

\usepackage{graphicx}	
\usepackage{amsmath}	
\usepackage{amssymb}	

\usepackage{hyperref}

\title[Polarisation in molecular clouds]{SILCC-Zoom: Polarisation and depolarisation in molecular clouds}
  \author[D. Seifried et al.]
  {D.~Seifried,$^1$\thanks{seifried@ph1.uni-koeln.de} S.~Walch,$^1$ S. Reissl,$^2$ J.~C.~Ib{\'a}{\~n}ez-Mej{\'i}a$^1$ \\
  $^1$Universit\"at zu K\"oln, I. Physikalisches Institut, Z\"ulpicher Str. 77, 50937 K\"oln, Germany\\
  $^2$Universit\"{a}t Heidelberg, Zentrum f\"{u}r Astronomie, Institut f\"{u}r Theoretische Astrophysik, Albert-Ueberle-Str. 2, 69120 Heidelberg, Germany\\
  }

\date{Released 2018}

\pagerange{\pageref{firstpage}--\pageref{lastpage}} \pubyear{2018}

\begin{document}

\label{firstpage}

\maketitle

\begin{abstract}
We present synthetic dust polarisation maps of 3D magneto-hydrodynamical simulations of molecular clouds before the onset of stellar feedback. The clouds are modelled within the SILCC-Zoom project and are embedded in their galactic environment. The radiative transfer is carried out with POLARIS for wavelengths from 70 $\mu$m to 3 mm at a resolution of 0.12 pc, and includes self-consistently calculated alignment efficiencies for radiative torque alignment. We explore the reason of the observed depolarisation in the center of molecular clouds: We find that dust grains remain well aligned even at high densities ($n$ $>$ 10$^3$ cm$^{-3}$) and visual extinctions ($A_\rmn{V}$ $>$ 1). The depolarisation is rather caused by strong variations of the magnetic field direction along the LOS due to turbulent motions. The observed magnetic field structure thus resembles best the mass-weighted, line-of-sight averaged field structure. Furthermore, it differs by only a few 1$^\circ$ for different wavelengths and is little affected by the spatial resolution of the synthetic observations. Noise effects can be reduced by convolving the image. Doing so, for $\lambda$ $\gtrsim$ 160 $\mu$m the observed magnetic field traces reliably the underlying field in regions with intensities $I$ $\gtrsim$ 2 times the noise level and column densities above 1 M$_{\sun}$ pc$^{-2}$. Here, typical deviations are \mbox{$\lesssim$ 10$^\circ$}. The observed structure is less reliable in regions with low polarisation degrees and possibly in regions with large column density gradients. Finally, we show that a simplified and widely used method without self-consistent dust alignment efficiencies can provide a good representation of the observable polarisation structure with deviations below 5$^\circ$.
\end{abstract}

\begin{keywords}
 MHD -- radiative transfer -- methods: numerical -- techniques: polarimetric -- ISM: clouds -- ISM: magnetic fields
\end{keywords}



\section{Introduction}

Magnetic fields seem to play an important role in the evolution of gas in disc galaxies, from the diffuse interstellar medium \citep[ISM; see e.g. the reviews by][]{Crutcher12,Beck13} to dense molecular clouds (MCs) and star forming cores. However, magnetic fields are observed only indirectly e.g. via polarised radiation emitted by dust grains. Both, the polarisation degree and polarisation angle reveal important information about the field strength and orientation. In particular with the advent of the BlastPol experiment \citep{Matthews14,Fissel16,Fissel18,Gandilo16,Santos17,Soler17,Ashton18} and the Planck satellite \citep{PlanckXX,PlanckXXII}, more and more dust polarisation observations for MCs have been reported in the literature \citep[see also e.g.][]{Houde04,Dotson10,Li13,Pillai15}. Also on smaller scales of individual protostellar cores, polarisation observations with ALMA become available \citep[e.g.][]{Hull17a,Hull17b,Koch18}. Their interpretation, however, and in particular the importance of magnetic fields on the cloud formation process is still under debate. Observations indicate a possible change of the orientation of the magnetic field from the diffuse, atomic to the dense molecular structures, which could indicate that magnetic fields act in guiding material to the dense central regions of MCs \citep[e.g.][]{Li13,Malinen16,Soler17,Fissel18,Monsch18}.

Furthermore, the polarisation degree is typically found to decrease towards the dense regions of MCs \citep[e.g.][]{Wolf03,Attard09,Tang09,Bertrang14,Brauer16,Fissel16,Santos17,Galametz18,Koch18,Soam18}. To date it is unclear what causes this depolarisation \citep[see e.g. the review by][]{Li14}. Possible reasons discussed in the literature are a reduced alignment efficiency of dust grains due to shielding of the interstellar radiation field, strong variations of the magnetic field direction along the line-of-sight, and resolution effects.

For these reasons it is essential to produce fully self-consistent synthetic dust polarisation maps from magneto-hydrodynamical (MHD) simulations, which help to interpret observational results. Such theoretical predictions, however, are challenging for three reasons: First, self-consistent MHD simulations have to be performed, capturing the thermal, dynamical, and chemical complexity of the ISM. Second, accurate theoretical models to calculate dust alignment efficiencies have to be applied \citep[see e.g.][for reviews]{Lazarian07review,Andersson15}. Third, full radiative transfer calculations are necessary to produce synthetic dust polarisation maps.

To date, a number of synthetic dust polarisation observations produced from simulations of MCs have been presented in the literature \citep[e.g.][]{Heitsch01,Padoan01,Pelkonen07,Pelkonen09,Kataoka12,Soler13,PlanckXX,King18,Vaisala18}. Despite the large number of works, they usually lack at least one of the aforementioned requirements, and thus do not produce \textit{fully} self-consistent dust polarisation maps.

Numerical simulations often omit the self-consistent treatment of the larger-scale galactic environment in which the clouds are embedded and rather consider the evolution of isolated MCs \citep[e.g.][]{Bate09,Clark11,Ward14,Iffrig15,Szucs16}. However, due to magnetic flux-freezing, the magnetic field pervading the MC is anchored to the ambient medium, which makes it inevitably necessary to model the ambient medium \textit{simultaneously} with the evolution of the cloud. Starting with an already existing cloud imposes the difficulty of realistic initial conditions \citep{Rey15}, which becomes even more important in the presence of magnetic fields. Hence, for self-consistent MC simulations, the formation history of the MC has to be modelled as well. These requirements in combination with the necessity for a high spatial resolution \citep[$\leq$ 0.1 pc,][]{Seifried17} impose significant computational demands, which make fully self-consistent simulations of MC formation and evolution challenging.

Furthermore, the works on synthetic dust polarisation observations mentioned before have not (yet) combined a self-consistent calculation of the grain alignment efficiencies with a subsequent radiative transfer to calculate all Stokes parameters, but are usually lacking one or both of these two aspects. First attempts to combine dust grain alignment theories with simplified MHD simulations were presented by \citet{Bethell07} and \citet{Pelkonen07,Pelkonen09}. The most frequently used approach so far, however, is to calculate (approximated) Stokes parameters based on the approximations given by \citet{Lee85}, \citet{Wardle90}, and \citet{Fiege00}. Since this approach does not consider any wavelength dependence and only polarisation due to thermal re-emission, it cannot make predictions for the polarisation of background radiation due to dichroic extinction \citep[see e.g.][for recent observations]{Li13,Hoq17}. Moreover, a general proof-of-concept of this approach by comparing with a self-consistent method is still lacking.

Recently, \citet{Reissl16} have presented the freely available dust polarisation radiative transfer code POLARIS, which is able to calculate grain alignment efficiencies and the subsequent radiative transfer in a fully self-consistent manner. We have already applied the code to the case of protostellar outflows showing the necessity of wavelength-dependent radiative transfer \citep{Reissl17} and filaments showing the advantage of additional Zeeman observations \citep{Reissl18}.

Using POLARIS and the SILCC-Zoom simulations of \textit{pristine} MCs, i.e. MCs before the onset of stellar feedback, evolving in their large-scale galactic environment \citep{Seifried17}, we try to remedy the shortcomings discussed above. We apply the POLARIS code to the simulation in a post-processing step to create synthetic dust polarisation maps, which can be used to interpret actual observations and to make predictions for future observations.

The structure of the paper is as follows: First, we present the initial conditions and numerical methods used for the MHD simulations (Section~\ref{sec:numerics}) and the subsequent radiative transfer with POLARIS (Section~\ref{sec:polaris}). We present our results in Section~\ref{sec:results} and discuss the wavelength dependence as well as the effect of noise on the observable polarisation structure. In Section~\ref{sec:correlation} we develop guidelines for actual observations in order to assess in which regions the observed structure is consistent with the underlying magnetic field structure. In Section~\ref{sec:depolarisation}, we investigate the cause of the  frequently observed depolarisation in MCs. Finally, we discuss our results and compare them with those using a simplified method in Section~\ref{sec:discussion} before we conclude in Section~\ref{sec:conclusion}.

\section{Numerics and initial conditions}
\label{sec:numerics}

The simulations presented here make use of the SILCC-Zoom simulations \citep{Seifried17} of pristine MCs which are based on the multi-phase ISM simulations carried out in the SILCC project \citep{Walch15,Girichidis16}. In the following we briefly describe the numerical methods used; for more details we refer the reader to the aforementioned papers.

The simulations are performed with the adaptive mesh refinement code FLASH version 4 \citep{Fryxell00,Dubey08}. We use an MHD solver which guarantees positive entropy and density \citep{Bouchut07,Waagan09}. We model the chemical evolution of the ISM using a simplified chemical network for H$^+$, H, H$_2$, C$^+$, CO, e$^-$, and O \citep{Nelson97,Glover07b,Glover10}, which also describes the thermal evolution of the gas including the most relevant heating and cooling processes. The shielding of the interstellar radiation field (ISRF) with $G_0$ = 1.7 \citep{Habing68,Draine78}, necessary for the chemical reactions and heating processes, is calculated according to the local column densities of total gas, H$_2$, and CO via the TreeCol algorithm \citep{Clark12b,Walch15,Wunsch17}. The dust temperature -- required for the dust radiative transfer -- is calculated self-consistently within the chemical network by assuming that the dust is in thermal equilibrium. The dust grains are heated by the ISRF and collision with the ambient gas particles and cool by thermal emission in the infrared, which is assumed to be optically thin \citep[see Section 2.2.6 in][for details]{Walch15}. We solve the Poisson equation for the self-gravity with a tree based method \citep{Wunsch17} and furthermore include a background potential due to the pre-existing stellar component in the galactic disc, which is modelled with an isothermal sheet with \mbox{$\Sigma_\rmn{star}$ = 30 M$_{\sun}$ pc$^{-2}$} and a scale height of 100 pc \citep{Spitzer42}.

The initial conditions are chosen to represent a section of a galactic disc with solar neighbourhood properties. The simulation box has a size of 500 pc $\times$ 500 pc $\times$ $\pm$ 5 kpc with periodic boundary conditions applied along the $x$- and $y$-direction and outflow conditions along the $z$-direction. The initial gas surface density is \mbox{$\Sigma_\rmn{gas}$ = 10 M$_{\sun}$ pc$^{-2}$} and the vertical distribution of the gas is modelled with a Gaussian profile
\begin{equation}
 \rho(z) = \rho_0 \times \textrm{exp}\left[ - \frac{1}{2} \left( \frac{z}{h_z} \right)^2 \right]
\end{equation}
with $h_z$ = 30 pc and $\rho_0 = 9 \times 10^{-24}$ g cm$^{-3}$. The gas near the disc midplane has an initial temperature of 4500 K and consists of atomic hydrogen and C$^+$. We initialize the magnetic field along the $x$-direction as
\begin{equation}
 B_{x} = B_{x,0} \sqrt{\rho(z)/\rho_0} \, ,
\end{equation}
where we set the magnetic field in the midplane to \mbox{$B_{x,0}$ = 3 $\mu$G} in accordance with recent observations \citep[e.g.][]{Beck13}.

From the start we inject supernovae (SNe) up to $t_0$ with a constant rate of 15 SNe Myr$^{-1}$. This SN rate is in agreement with the Kennicutt-Schmidt star formation rate surface density at the given gas surface density \citep{Schmidt59,Kennicutt98}. Half of the SNe are randomly placed in the $x$-$y$-plane following a Gaussian profile with a scale height of 50 pc in the vertical direction, the other half is placed at density peaks. This SN distribution allows us to obtain a realistic, turbulent, multiphase ISM as initial conditions for the subsequent zoom-in procedure \citep{Walch15,Girichidis16}. For a single SN explosion we inject 10$^{51}$ erg in the form of thermal energy if the Sedov-Taylor radius is resolved with at least 4 cells. Otherwise, we heat the gas within the injection region to $10^4$ K and inject the momentum, which the swept-up shell has gained at the end of the pressure-driven snowplough phase \citep[see][for details]{Gatto15}.

The grid resolution up to $t_0$ is 3.9 pc. At $t_0$ we stop the injection of further SNe and choose six different cuboid-like regions in which MCs are about to form. The six clouds -- henceforth denoted as MC1 to MC6 -- have partly different $t_0$ and differently sized cuboids (see Table~\ref{tab:overview}). Beginning at $t_0$, we continue the simulation for another 1.5 Myr over which we progressively increase the spatial resolution in these regions from 3.9 pc to 0.12 pc \citep[see Table 2 in][]{Seifried17}, whereas in the surroundings we keep the lower resolution of 3.9 pc. Afterwards we continue the simulations with the highest resolution of 0.12 pc in the zoom-in regions. We thus are able to resolve the filamentary structure of the clouds while simultaneously including the effect of accretion onto the clouds from the larger-scale, galactic environment.

\begin{table}
\caption{Overview of the simulations giving the run name, the starting time of the zoom-in procedure $t_0$, the time $t_\rmn{end}$ over which the clouds are evolved after $t_0$, the extent of the cuboid-like zoom-in region, and its center in the simulation domain.}
\centering
\begin{tabular}{ccccc}
  \hline
 run & $t_0$ & $t_\rmn{end}$  & volume & center \\
  & (Myr) & (Myr) & pc$^3$ & (pc) \\
 \hline
 MC1 & 16.0 & 5.5 & 130 $\times$ 110 $\times$ 110 & (-84, 100, 0) \\
 MC2 & 16.0 & 5.5 & 130 $\times$ 110 $\times$ 104 & (126, 117, 0)  \\
 MC3 & 16.0 & 3.0 & 120 $\times$ 110 $\times$ 104 & (-125, -103, 0)  \\
 MC4 & 11.6 & 3.0 & 117 $\times$ 130 $\times$ 97 & (-97, 126, 0)  \\
 MC5 & 11.6 & 3.0 & 91 $\times$ 97 $\times$ 97 & (3, 16, 0)  \\
 MC6 & 16.0 & 3.0 &  130 $\times$ 136 $\times$ 104 & (62, 175, 0)  \\
 \hline
 \end{tabular}
 \label{tab:overview}
\end{table}

\section{Radiative transfer}
\label{sec:polaris}

The radiative transfer (RT) calculations are performed with the freely available RT code POLARIS \footnote{http://www1.astrophysik.uni-kiel.de/$\sim$polaris} \citep{Reissl16}. POLARIS is a newly designed 3D dust continuum Monte-Carlo code allowing to solve the RT problem including dust polarisation on an octree grid. The octree structure agrees with the FLASH file format, making it possible to post-process MHD data without interpolating it on a regular grid.  We have already successfully applied the code to simulations of protostellar outflows \citep{Reissl17}. In the following we give a short summary of the applied techniques; for further details see \citet{Reissl16,Reissl17}. The major parameters are summarised in Table~\ref{tab:parameters}.
\begin{table*}
\caption{Overview of the most important parameters used in the dust polarisation radiative transfer.}
\centering
\begin{tabular}{cc}
  \hline
 parameter & description \\
 \hline
 $\delta_0$ & upper size threshold for IDG, up to which grains are aligned with the magnetic field (Eq.~\ref{eq:delta}) \\
 $a_\rmn{alig}$ &  lower size threshold for RAT, above which grains are aligned with the magnetic field (Eq.~\ref{eq:omega}) \\
 $a_\rmn{l}$  & upper size threshold for RAT, up to which grains are aligned with the magnetic field (Eq.~\ref{eq:a_l}) \\ 
 $a_\rmn{min}$ & minimum grain size in the dust model used, set to 5~nm \\
 $a_\rmn{max}$ & maximum grain size in the dust model used, set to 2~$\mu$m \\
 \hline
 \end{tabular}
 \label{tab:parameters}
\end{table*}

\subsection{Imperfect Davis-Greenstein alignment}

As noted before, the major obstacle in dust polarisation RT simulations is the lack of a coherent dust grain alignment theory. For this reason, POLARIS supports all major classes of grain alignment theories that have been proposed over the years \citep[see e.g.][for reviews]{Lazarian07review,Andersson15}.

The imperfect Davis-Greenstein alignment (IDG) theory is the classic theory to account for dust grains to align with their shorter axis along the magnetic field \citep{Davis51}. Here, dust grains acquire angular momentum by random gas collisions and align with the magnetic field by means of dissipation of the angular momentum components perpendicular to the magnetic field. Balancing the characteristic time scales results in the characteristic quantity of
\begin{equation}
        \delta_0 = 2.06  \left(\frac{B}{\rmn{1~\mu G}}\right)^2 \left(\frac{n_\rmn{g}}{1~\rmn{cm}^{-3}} \, \frac{T_\rmn{d}}{\rmn{1~K}} \, \sqrt{\frac{T_\rmn{g}}{\rmn{1~K}}}\right)^{-1} \, \mu\rmn{m} \, ,
\label{eq:delta}
\end{equation}
where $n_\rmn{g}$ is the gas number density, $B$ the magnetic field strength, and $T_\rmn{g}$ and $T_\rmn{d}$ are the gas and dust temperature, respectively. The quantity $\delta_0$ is an upper threshold for the dust grain size, since grains larger that $\delta_0$ do not significantly contribute to linear polarisation.

When using IDG for the RT calculations, we find that the IDG mechanism is highly inefficient in aligning the dust grains with the magnetic field. The obtained polarisation degree is below 1\%, and thus one order of magnitude lower than that for radiative torque alignment (see below). Since such small polarisation fractions are not measurable in observations, we neglected IDG in this work. We note that this result is in good agreement with our findings for much denser environments of protostellar outflows, where IDG does not contribute measurably to the polarisation degree either \citep{Reissl17}.

\subsection{Radiative torque alignment}
\label{sec:RAT}

The radiative torque (RAT) alignment theory matches most of the observational constraints \citep{Lazarian07,Andersson15}. Left and right circularly polarized, unidirectional light acts differently on grains with a net helicity resulting in a net torque spinning up the dust grain. In order to overcome randomization, the angular frequency due to radiation, $\omega_\rmn{rad}$, has to be larger than that caused by random gas bombardment, $\omega_\rmn{gas}$, by about a factor of $\sim$ 3 \citep{Hoang08}. Following \citet{Bethell07}, this ratio can be phenomenologically described by
\begin{equation}
 \left(\frac{\omega_{\rmn{rad}}}{\omega_{\rmn{gas}}}\right)^2 \propto a_\rmn{alig}\left(\frac{1}{n_{\rmn{g}}k_\rmn{B} T_{\rmn{g}}} \int Q_{\Gamma}\lambda u_{\rmn{\lambda}} d\lambda \right)^2 \, ,
\label{eq:omega}
\end{equation}
where $Q_{\Gamma}$ is the alignment efficiency depending on the grain shape and the wavelength $\lambda$, $u_\rmn{\lambda}$ is the energy density of the local radiation field, and $k_\rmn{B}$ the Boltzmann constant. Here, $a_{\rm{alig}}$ is a lower threshold and dust grains smaller than $a_\rmn{alig}$ are not aligned with the magnetic field, i.e. do not contribute to the polarised emission. We use a standard ISRF taken from to \citet{Moskalenko06} to calculate $a_\rmn{alig}$.

A rotating dust grain then gains a net magnetization via the Barnett effect \citep{Barnett17} leading to subsequent alignment with the magnetic field. The alignment efficiency is governed by \citep{Lazarian07review,Hughes09}
\begin{equation}
 a_{\rm l} =  2.44 \times 10^8 \, s^2 \, \left(\frac{B}{\rmn{1~\mu G}}\right)  \left(\frac{n_\rmn{g}}{1~\rmn{cm}^{-3}}  \frac{T_\rmn{d}}{\rmn{1~K}} \, \sqrt{\frac{T_\rmn{g}}{\rmn{1~K}}}\right)^{-1} \,  \mu\rmn{m} \, ,
\label{eq:a_l}
\end{equation}
where $ a_\rmn{l}$ is an upper size threshold and $s$ the aspect ratio of the grains. Dust grains larger than $a_\rmn{l}$ cannot efficiently align with the magnetic field within a Larmor precession time. We note that for the conditions typical for MCs discussed here, $a_\rmn{l}$ is always above 2 $\mu$m, which is the upper size limit of our applied dust grain model, and thus has no effect on the polarisation.

\subsection{Radiative transfer}

In this paper, we focus on conditions typical for the ISM. Hence, we apply a dust model consisting of 37.5\% graphite and 62.5\% amorphous silicate grains that reproduces the extinction curve of our own Galaxy \citep{Mathis77}. The dust density is obtained from the gas density assuming a spatially constant dust-to-gas mass ratio of 1\%. The grains are assumed to have a size distribution of $n(a) \propto a^{-3.5}$ with the canonical values of the lower and upper cut-off radius of $a_{\rm min}$ = 5 nm and \mbox{$a_{\rm max}$ = 2 $\mu$m}, respectively. The shape of a single dust grain is fractal in nature. However, we apply an oblate shape with an aspect ratio of $s = 0.5$, a valid approximation for an averaged ensemble of dust grains \citep{Hildebrand95,Draine17}. We pre-calculate individual cross sections for 160 size bins and 104 wavelength bins \citep[see][for details]{Reissl17} with the scattering code DDSCAT \citep{Draine13}. For the optical properties of the different materials we take the tabulated data from \cite{Lee85} and \citet{Laor93}.

We perform dust RT calculations for wavelengths of \mbox{$\lambda$ = 70.4,} 161, 243, 342, 515, 850, 1300, and \mbox{3000 $\mu$m}. The resolution of the obtained dust emission maps is identical to that of the highest resolution in the MHD simulations of 0.12 pc. Assuming a distance of the clouds of 1 kpc, this corresponds to a resolution of 25'', i.e. comparable to that of the BlastPol experiment with a resolution of 30'', 42'', and 60'' at wavelengths of 250, 350, and 500 $\mu$m, respectively \citep{Galitzki14,Fissel16,Fissel18,Gandilo16}. We also discuss the effect of lowering the resolution of the synthetic observations by a factor of 2, 4, 8, and 16 in Appendix~\ref{sec:resolution}.

From the RT calculations we obtain the Stokes parameters $I$, $Q$, and $U$, where $I$ is the total intensity, and $Q$ and $U$ quantify the linear polarisation of the observed radiation. The polarisation angle $\varphi$ is calculated as
\begin{equation}
 \varphi = \frac{1}{2}\rmn{arctan}(U,Q  ) \,
 \label{eq:phi}
\end{equation}
and the polarisation degree is given by 
\begin{equation}
 p = \frac{\sqrt{Q^2 + U^2}}{I} \, .
 \label{eq:p}
\end{equation}

\section{Results}
\label{sec:results}

\subsection{Simulation overview}
\label{sec:overview}

\begin{figure}
 \includegraphics[width=\linewidth]{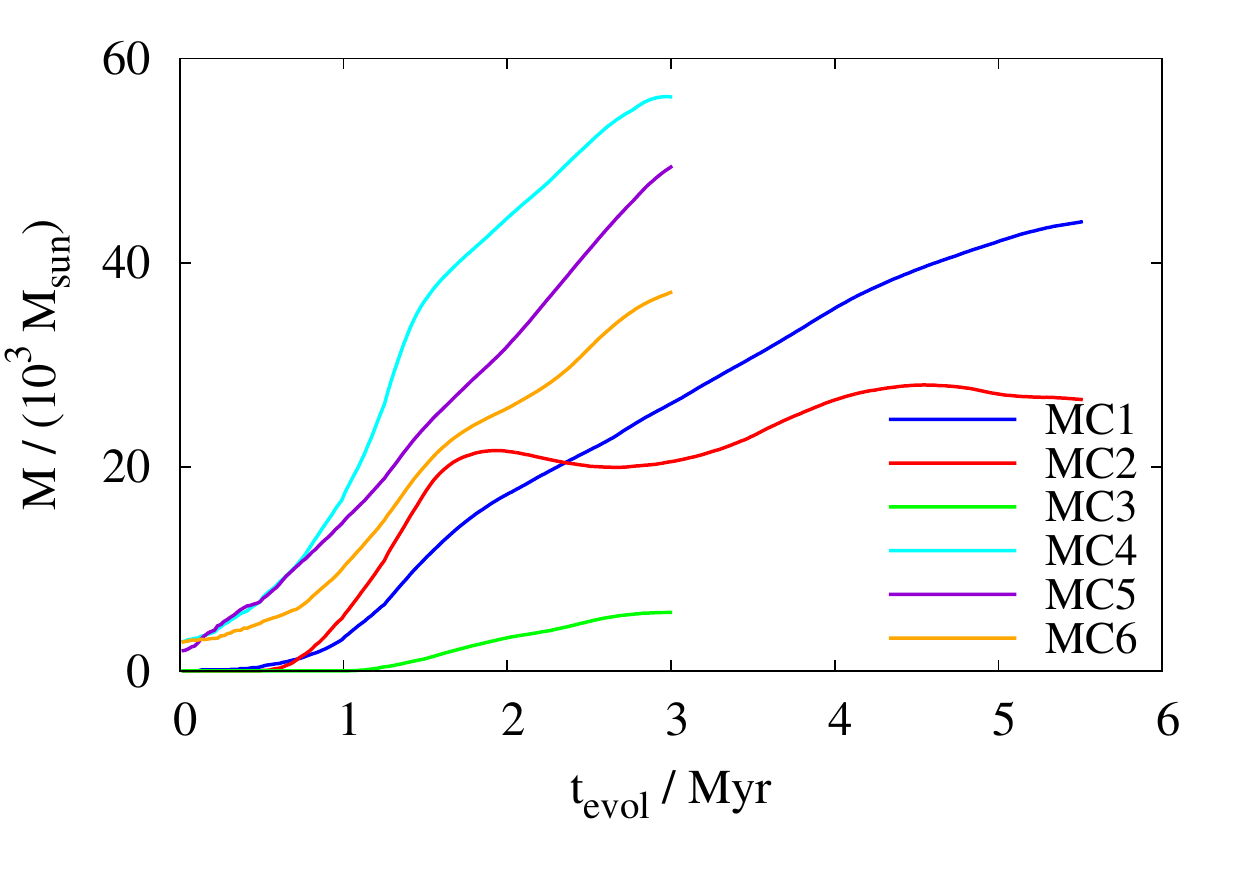}
 \caption{Time evolution of the masses of the six MCs defined as all gas with number densities above 100 cm$^{-3}$. Most of the results in this paper are presented for $t_\rmn{evol}$ = 3 Myr measured from $t_0$ onwards.}
 \label{fig:masses}
\end{figure}

In Table~\ref{tab:overview} we list basic properties of the six simulations. We start to zoom-in at $t_0$, at which the clouds start to condense out of the diffuse ISM. From $t_0$ on, we evolve  MC4 -- MC6 for additional 3 Myr and MC1 and MC2 for 5.5 Myr in order to test a potential long-term time dependence. Throughout the paper we refer to the time elapsed since $t_0$ as \mbox{$t_\rmn{evol}$ = $t$ - $t_0$.}

\begin{figure}
 \includegraphics[width=\linewidth]{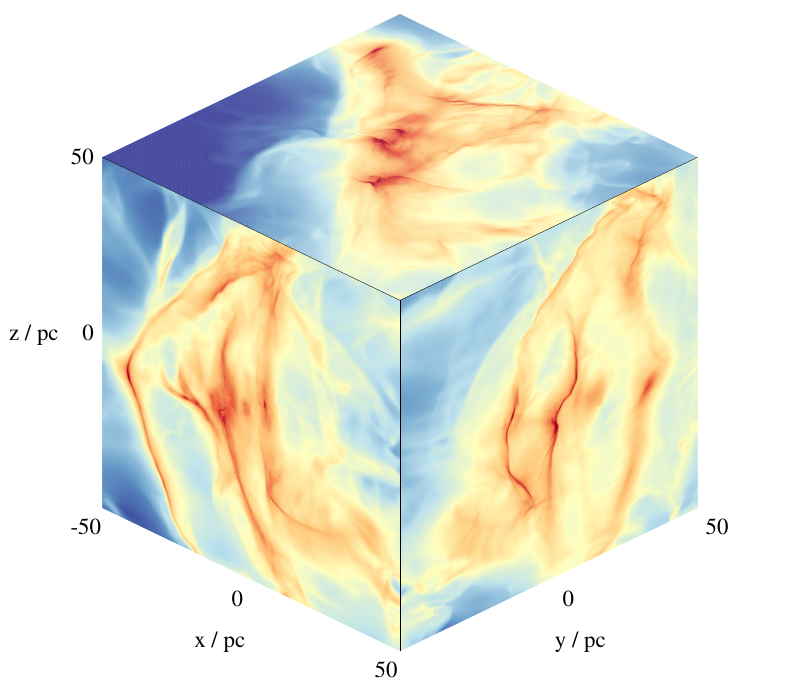}
 \caption{Column density of MC1 at $t_\rmn{evol}$ = 3 Myr from all three sides showing the filamentary structure of the cloud.}
 \label{fig:cd}
\end{figure}

In Fig.~\ref{fig:masses} we show the time evolution of the mass of all gas in the six zoom-in regions, which has a density above \mbox{3.84 $\times$ 10$^{-22}$ g cm$^{-3}$}, corresponding to a number density of 100 cm$^{-3}$ assuming a mean molecular weight of 2.3. The typical masses of the MCs are of the order of a few 10$^4$ M$_{\sun}$ and thus representative for Galactic MCs \citep{Larson81,Solomon87,Elmegreen96,Heyer01,Roman10,Miville17}. In order to obtain a visual impression of the 3D structure of the clouds, we show the column density projection of MC1 at  $t_\rmn{evol}$ = 3 Myr in Fig.~\ref{fig:cd}. The cloud shows a pronounced filamentary structure partly shaped by SNe going off during its earliest evolutionary phase (before $t_0$). Qualitatively similar results are also found for the other MCs. We will postpone the analysis of the influence of magnetic fields on the evolution of the clouds to subsequent work and only focus on the observable polarisation structure.

\subsection{Wavelength dependence}
\label{sec:lambda}

\begin{figure*}
\centering
 \includegraphics[width=\textwidth]{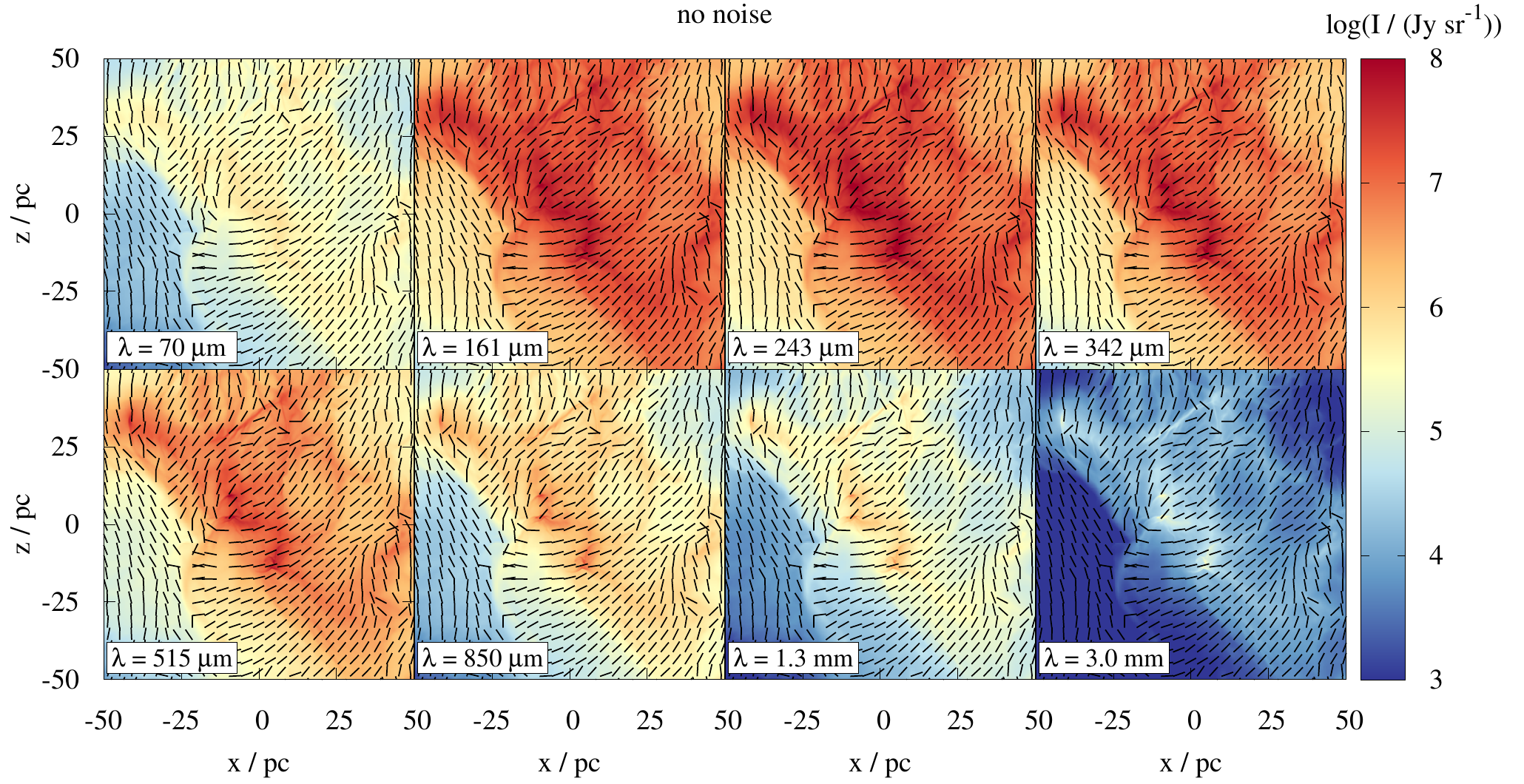}
 \caption{Dust emission intensity and polarisation vectors (black bars) for MC1 at $t_\rmn{evol}$ = 3 Myr along the $z$-direction for wavelengths from 70 $\mu$m to 3 mm (top left to bottom right). The polarisation pattern is remarkably similar for all $\lambda$. Since the polarisation is due to thermal re-emission of the dust particles, the magnetic field orientation can be obtained by rotating the vectors by 90$^\circ$.}
 \label{fig:maps}
\end{figure*}

\begin{figure}
 \centering
 \includegraphics[width=\linewidth]{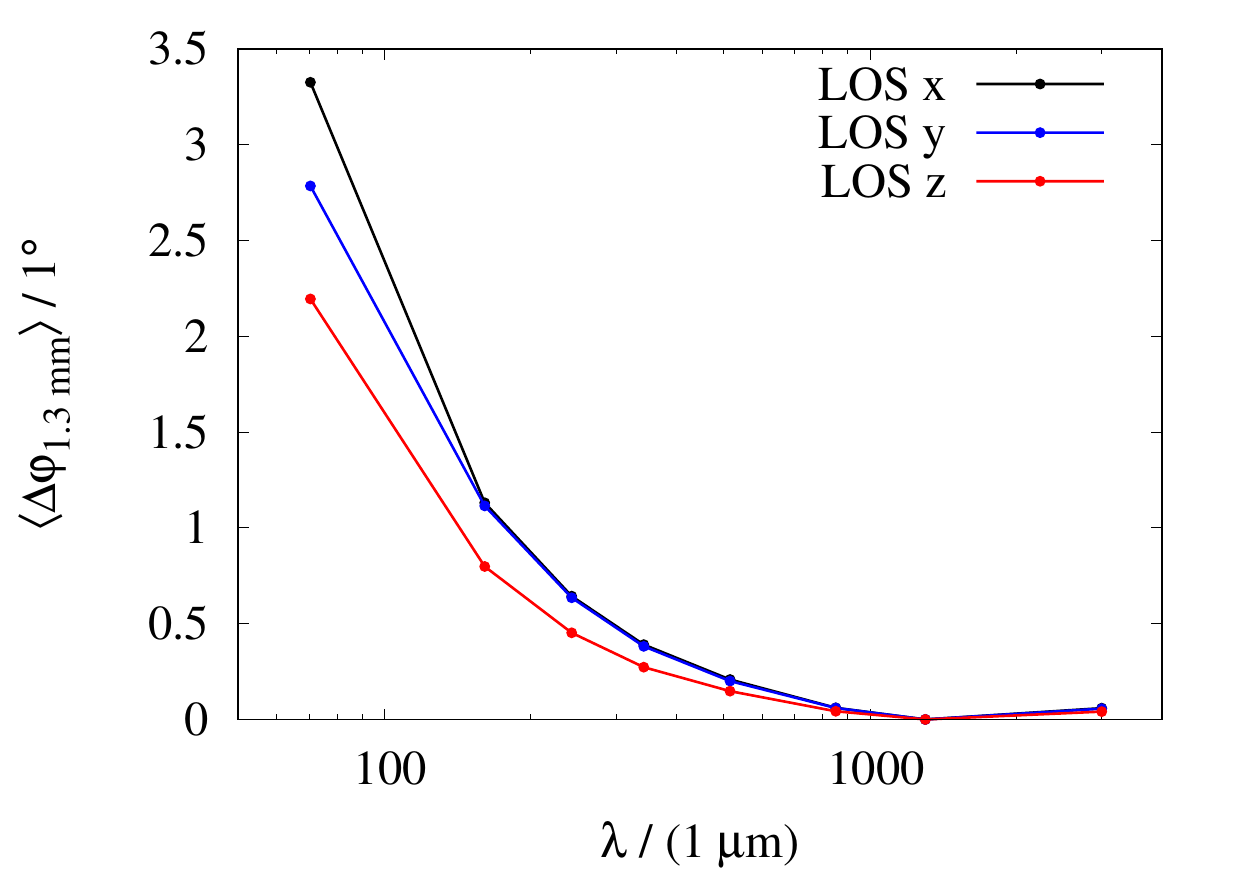}
 \caption{Mean difference $\left\langle \Delta \varphi_\rmn{1.3 mm} \right\rangle$ of the polarisation pattern for different wavelengths with respect to that of $\lambda_\rmn{ref}$ = 1.3 mm averaged over all pixels in the maps shown in Fig.~\ref{fig:maps} for MC1 at $t_\rmn{evol}$ = 3 Myr along three different directions.}
 \label{fig:dphi_lambda}
\end{figure}

The polarisation results discussed in the following span a four-dimensional parameter space: the wavelength, the line-of-sight (LOS, here the $x$-, $y$-, and $z$-direction), the time, and the different clouds. For this reason, we first consider the wavelength dependence of the observable polarisation pattern for MC1 at \mbox{$t_\rmn{evol}$ = 3 Myr} along the $z$-direction in Fig.~\ref{fig:maps}.

We find that the intensity, $I$, of the dust emission varies strongly with $\lambda$ as expected for a blackbody radiator at about 10~K. It peaks around 200 -- 400 $\mu$m and declines towards longer wavelengths; at $\lambda$ = 70 $\mu$m the emission is very weak, as expected for a cloud without any internal heating sources. Furthermore, the polarisation pattern is remarkably similar for all $\lambda$ considered. We attribute this to the fact that for $\lambda \geq$ 70 $\mu$m we are already in the regime of thermal re-emission, i.e. the transition from dichroic extinction to thermal re-emission occurs at shorter $\lambda$, similar to our results for protostellar outflows \citep{Reissl17}. Consequently, the observed polarisation is due to emission of the dust grains along their major axis and the polarisation vectors have to be rotated by 90$^\circ$ in order to obtain an impression of the magnetic field structure.

In order to quantify the difference in the polarisation pattern for the different $\lambda$, for each pixel in the emission map we calculate the difference $\Delta \varphi$ of the polarisation angle for a given wavelength, $\varphi_{\lambda}$ (Eq.~\ref{eq:phi}), to that of a reference wavelength, $\varphi_\rmn{\lambda_\rmn{ref}}$. We choose \mbox{$\lambda_\rmn{ref}$ = 1.3 mm}, frequently observed due to its correspondence to the CO(2--1) line transition, and calculate the residual between both angles as given by Eq.~7 in \citet{PlanckXIX} for two different sets of Stokes parameters
\begin{flalign}
 &\Delta \varphi_\rmn{1.3 mm}(\lambda) = \varphi_{\lambda} - \varphi_\rmn{1.3 mm} \, \nonumber \\
 &\equiv \frac{1}{2}\text{arctan}(Q_{\lambda} U_\rmn{1.3 mm} - U_{\lambda}  Q_\rmn{1.3 mm},\, Q_{\lambda}  Q_\rmn{1.3 mm} + U_{\lambda}  U_\rmn{1.3 mm}) \, .
\end{flalign}

In Fig.~\ref{fig:dphi_lambda}, we show the mean difference, $\left\langle \Delta \varphi_\rmn{1.3 mm} \right\rangle$, averaged over all pixels in the map for MC1 at $t_\rmn{evol}$ = 3 Myr along the three different LOS for the different wavelengths considered. The relative differences in $\varphi$ are almost negligible with values around 1$^\circ$ or below and decreases with increasing $\lambda$. Also for all other considered MCs and times, we find  very similar results. Our findings thus indicate that in actual observations the polarisation structure of MCs can be determined accurately even by single-wavelength observations when taken in the regime of thermal re-emission, which is supported by recent observations of \citet{Liu18}. Due to the small values of $\left\langle \Delta \varphi_\rmn{1.3 mm} \right\rangle$, in the remainder of the paper we consider \mbox{$\lambda$ = 1.3 mm} as a representative case.

\subsection{What regions are probed by dust polarisation observations?}
\label{sec:accuracy}

\begin{figure*}
 \centering
 \includegraphics[width=\linewidth]{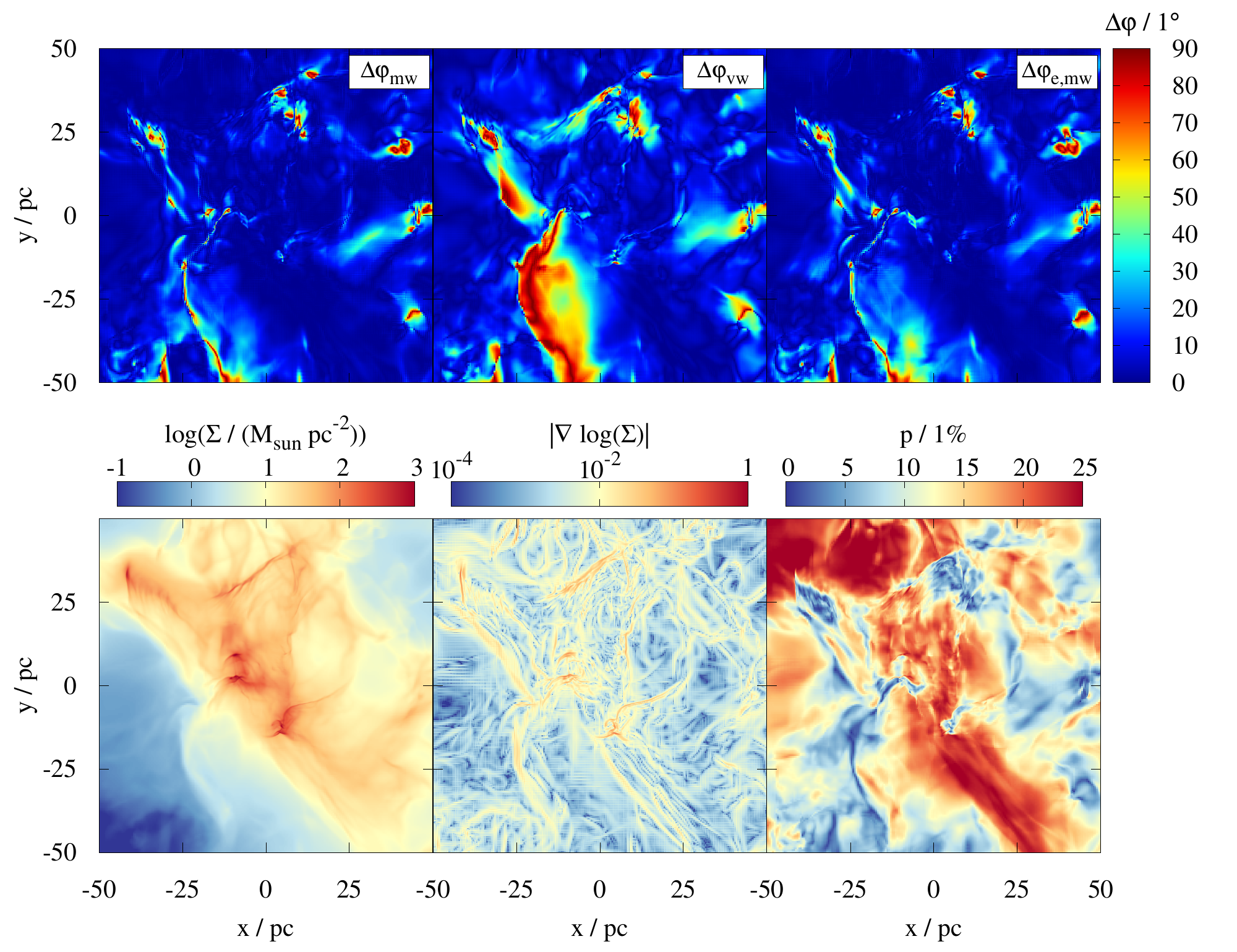}
 \caption{Top: Map of the deviation between the observed magnetic field (at $\lambda$ = 1.3 mm) and three different LOS-averages along the $z$-direction for MC1 at $t_\rmn{evol}$ = 3 Myr: the mass-weighted magnetic field, the volume-weighted magnetic field, and the mass-weighted magnetic field direction (from left to right, Eqs.~\ref{eq:Bvw} -- \ref{eq:eBmw}). The typical deviation is around 10$^\circ$, although regions with significant deviations up to 90$^\circ$ are present. Bottom: For comparison (see Section~\ref{sec:correlation}), we show the column density $\Sigma$, the gradient of log($\Sigma$), as well as the polarisation degree (Eq.~\ref{eq:p}). For ideal observations without noise, $\Delta \varphi$ is uncorrelated with both $\Sigma$ and \mbox{$|\nabla \textrm{log}(\Sigma)|$}, but seems to be anti-correlated with the polarisation degree.}
 \label{fig:dphi_map}
\end{figure*}

We would like to know what information about the actual magnetic field structure in the clouds can be extracted from dust polarisation observations. It is commonly assumed in the literature that the orientation of observed polarisation vectors is a good measure for the 2D, LOS-averaged magnetic field orientation in astronomical objects as both quantities present an average along a given LOS. However, despite this being an intuitive assumption, a strict proof-of-concept is lacking so far. Furthermore, we emphasise that the observed polarisation can only give an indication about the \textit{orientation}, not the \textit{direction} of the magnetic field. Hence, caution is required when inferring properties of the magnetic field from polarisation observations.

In this respect, the aim of this section is to investigate to which extent polarisation observations allow for conclusions about the 2D magnetic field structure (see also Section~\ref{sec:LOSvariations} for 3D effects). For this purpose, we compare the orientation of the polarisation vectors inferred from the synthetic dust polarisation maps with the magnetic field direction obtained directly from the simulation data. For the latter we apply three different ways of averaging along the LOS:
\begin{enumerate}
 \item the volume-weighted magnetic field
 \begin{equation}
  \mathbf{B}_\rmn{vw} = \frac{\int_\rmn{LOS} \, \mathbf{B} \, \rmn{d}l}{\int_\rmn{LOS} \, \rmn{d}l} \, ,
  \label{eq:Bvw}
 \end{equation}
\item the mass-weighted magnetic field
 \begin{equation}
  \mathbf{B}_\rmn{mw} = \frac{\int_\rmn{LOS} \, \rho \, \mathbf{B} \, \rmn{d}l}{\int_\rmn{LOS} \, \rho \, \rmn{d}l} \, ,
  \label{eq:Bmw}
 \end{equation}
 \item and the mass-weighted magnetic field {\it direction}
 \begin{equation}
   \mathbf{e}_\rmn{B,mw} = \frac{\int_\rmn{LOS} \, \rho \, \mathbf{e}_\rmn{B} \, \rmn{d}l}{\int_\rmn{LOS} \, \rho \, \rmn{d}l} \;\; \textrm{with} \; \mathbf{e}_\rmn{B} = \frac{\mathbf{B}}{|\mathbf{B}|} \, .
   \label{eq:eBmw}
 \end{equation}
\end{enumerate}
In the definitions above we only consider the components perpendicular to the LOS, i.e. the vectors are projected to the plane-of-sky. That allows us to directly compare them with the obtained polarisation vectors. We note that we include $\mathbf{e}_\rmn{B,mw}$ as it does not depend on the magnetic field strength, similar as the polarised emission would do in the case of perfectly aligned dust grains.

We again emphasise that taking into account the actual direction of the magnetic field differs from the mechanism resulting in polarisation, where only the orientation plays a role: two anti-parallel vectors $\mathbf{B}_2$ = $-\mathbf{B}_1$ would result in a net polarisation, but would cancel out in Eqs.~\ref{eq:Bvw} -- \ref{eq:eBmw}. Nonetheless, the definitions given in the Eqs.~\ref{eq:Bvw} -- \ref{eq:eBmw} provide an intuitive measure of the average magnetic field direction in the plane-of-sky.

For the sake of simplicity, throughout the paper we denote the polarisation vectors rotated by 90$^\circ$ as the \textit{observed magnetic field} $\mathbf{B}_\rmn{\lambda}$ at a given wavelength $\lambda$. For \mbox{$\lambda$ = 1.3 mm} we now calculate for each pixel the angle difference between the orientation of the observed magnetic field $\mathbf{B}_\rmn{1.3mm}$ and one of the three LOS-averages defined above:
\begin{eqnarray}
  \Delta \varphi_\rmn{vw} &=& \measuredangle (\mathbf{B}_\rmn{1.3mm}, \, \mathbf{B}_\rmn{vw}) \, , \label{eq:delta_vw} \\
  \Delta \varphi_\rmn{mw} &=& \measuredangle (\mathbf{B}_\rmn{1.3mm}, \, \mathbf{B}_\rmn{mw}) \, , \label{eq:delta_mw} \\
  \Delta \varphi_\rmn{e,mw} &=& \measuredangle (\mathbf{B}_\rmn{1.3mm}, \, \mathbf{e}_\rmn{B,mw}) \, \label{eq:delta_emw} .
\end{eqnarray}
As the polarisation only contains information about the orientation, not the actual direction, both vectors are mapped on a common quadrant, i.e., the angle difference cannot exceed 90$^\circ$.

\begin{figure*}
 \centering
 \includegraphics[width=\linewidth]{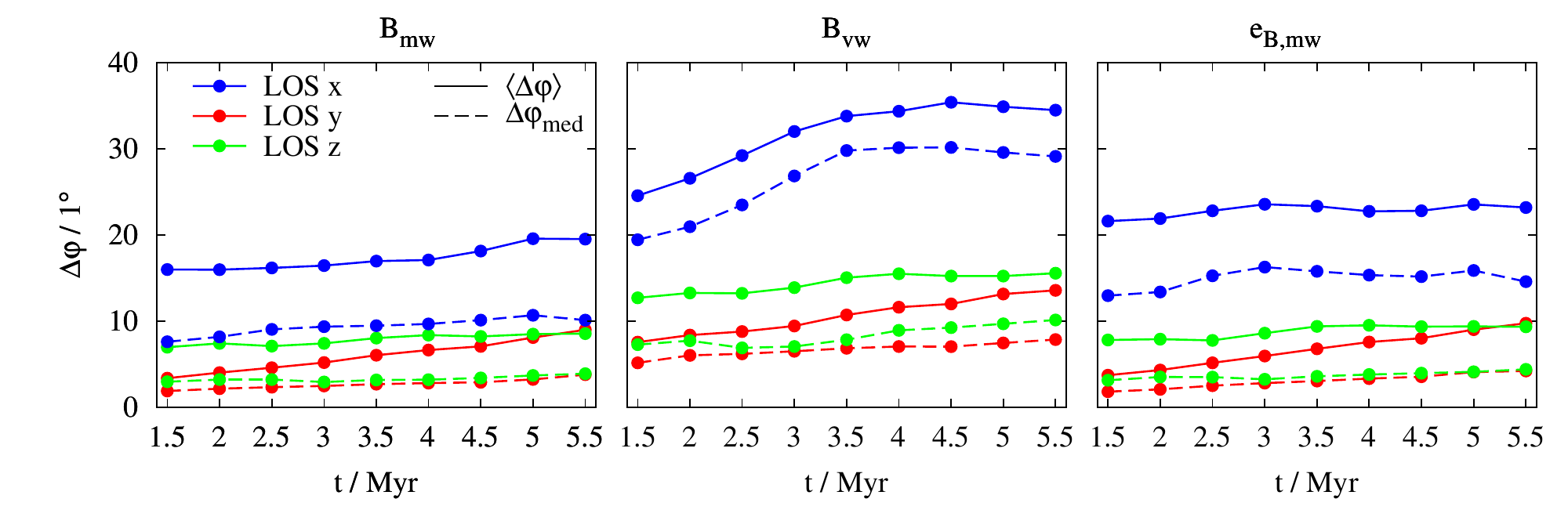}
 \caption{Mean (solid lines) and median deviation (dashed lines) between the observed magnetic field orientation at 1.3 mm and that of the mass- (left, Eq.~\ref{eq:delta_mw}) and volume-weighted LOS-averaged magnetic field (middle, Eq.~\ref{eq:delta_vw}) as well as the mass-weighted field direction (right, Eq.~\ref{eq:delta_emw}). The values are shown for MC1 for different directions (colour-coded). For all three definitions, $\Delta \varphi_\rmn{med}$ is about a factor of 1.5 -- 2 smaller than $\left\langle \Delta \varphi \right\rangle$. The deviation is largest along the direction of the initial magnetic field in the simulation setup ($x$-direction, blue curves). The observed field probes best the mass-weighted field (left).}
 \label{fig:dphi_mean}
\end{figure*}
In Fig.~\ref{fig:dphi_map} we show the map of $\Delta \varphi_\rmn{mw}$, $\Delta \varphi_\rmn{vw}$, and $\Delta \varphi_\rmn{e,mw}$ for MC1 along the $z$-direction at $t_\rmn{evol}$ = 3 Myr. For the majority of the pixels the typical deviations are of the order of 10$^\circ$ - 20$^\circ$ or lower. There are, however, a few regions where the observed and LOS-averaged magnetic field structure do not match well and show deviations of up to 90$^\circ$. Furthermore, $\Delta \varphi_\rmn{vw}$ seems to be somewhat larger than $\Delta \varphi_\rmn{mw}$ and $\Delta \varphi_\rmn{e,mw}$.

We calculate the (pixel-weighted) mean, $\left\langle \Delta \varphi \right\rangle$, as well as the median of the deviation, $\Delta \varphi_\rmn{med}$ in the entire map. We do this for the three different directions and different times from $t_\rmn{evol}$ = 1.5 Myr to 5.5 Myr (Fig.~\ref{fig:dphi_mean}). The results for MC1 show four characteristic features, which are also found for the other clouds.

First, $\Delta \varphi_\rmn{med}$ is in general about 1.5 -- 2 times smaller than $\left\langle \Delta \varphi \right\rangle$. We attribute this to the small regions with relatively high $\Delta \varphi$ ($\gtrsim$ a few 10$^\circ$, see Fig.~\ref{fig:dphi_map}), which increase the value of $\left\langle \Delta \varphi \right\rangle$. Hence, the median value seems to give a better impression of how well the observed field structure represents the LOS-averaged field structure.

Second, the deviations are largest along the $x$-direction, i.e. that of the magnetic field at the start of the simulations (blue curves). Here, the observable component of the magnetic field is the random component created by turbulent motions and the gravitational collapse. For the other two LOS we see the initial magnetic field which is reshaped during the formation of the cloud. It is thus not surprising that for a randomly oriented field the differences in $\mathbf{B}_\rmn{1.3mm}$ and the LOS-averaged magnetic field are larger than for a more ordered field.

Third, we find that the deviations show only little variation over time, in particular for the mass-weighted cases (left and right panel). Typical deviations increase over time by a few 10\% only.

Fourth, our results indicate that $\mathbf{B}_\rmn{1.3mm}$ probes best the mass-weighted magnetic field (left panel). The agreement of $\mathbf{B}_\rmn{1.3mm}$ with $\mathbf{e}_\rmn{B,mw}$ (right panel) is comparable to that with $\mathbf{B}_\rmn{mw}$, although, in particular along the $x$-direction, the deviations are a few 10\% higher. The largest deviations are found between $\mathbf{B}_\rmn{1.3mm}$ and the volume-weighted magnetic field (middle panel), with $\left\langle \Delta \varphi_\rmn{vw} \right\rangle$ being a few 10\% up to 2 times larger than $\left\langle \Delta \varphi_\rmn{mw} \right\rangle$. However, $\left\langle \Delta \varphi_\rmn{vw} \right\rangle$ remains smaller than an  deviation of 45$^\circ$ expected for a randomly distributed field.

To summarise, dust polarisation observations seem to probe the magnetic field structure in the dense regions along the LOS and less the lower-density regions in the fore- and background, which is why in the following we focus on the comparison between the observed magnetic field structure and $\mathbf{B}_\rmn{mw}$. We further emphasise that, although we could confirm a good match between the observed and the LOS-averaged magnetic field structure in pristine MCs, two restrictions have to be kept in mind: i) The polarisation only allows for conclusions about the 2D, average field orientation, but neither its direction nor the actual 3D structure (see also Section~\ref{sec:LOSvariations}) and ii) the result has to be confirmed independently under different conditions e.g. in the presence of stellar feedback (see Section~\ref{sec:caveats}) or on different scales like protostellar or galactic disks.

\subsection{The influence of noise}
\label{sec:noise}

\begin{figure}
 \includegraphics[width=\linewidth]{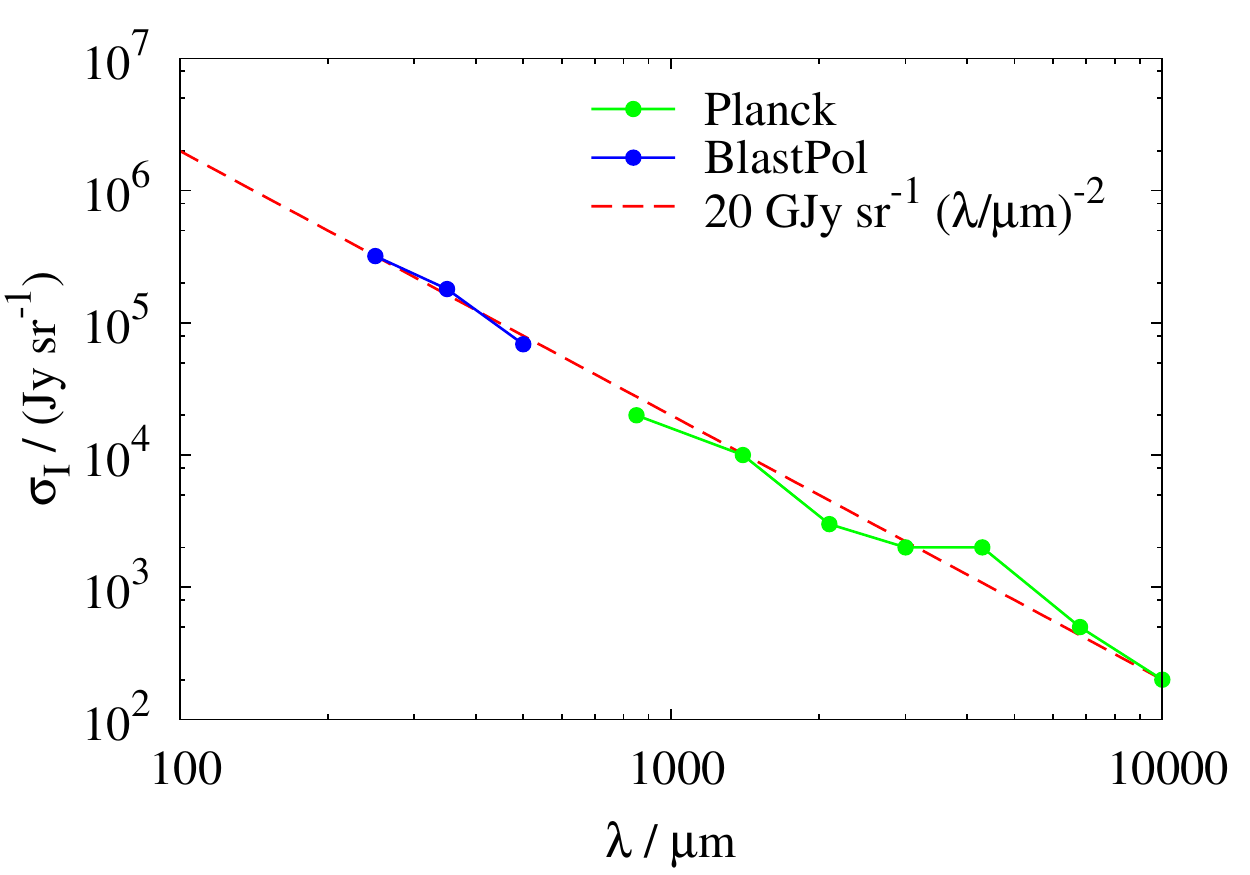}
 \caption{Wavelength dependence of the noise estimate for the BlastPol experiment \citep[blue line,][]{Fissel13} and the Planck satellite \citep[green line,][]{Pelkonen07}. The red dashed curve shows the noise approximation used in this work.}
 \label{fig:noise}
\end{figure}

In order to account for  the effect of noise on the polarisation structure, we add Gaussian distributed, white noise to the Stokes parameters. For each, $I$, $Q$, and $U$ we use \textit{independent} white noise at the same level. We are aware that this might represent some kind of simplification since e.g. the noise for $Q$ and $U$ is not necessarily Gaussian \citep{Wardle74,Montier15}. However, this approach probably represent the worst-case-scenario, which is why we expect the results presented in the following to hold also for more realistic noise models.

In order to estimate the noise level, $\sigma_I$ (also used for $Q$ and $U$), we take typical observational noise estimates of the BlastPol experiment \citep[][Table 3.1\footnote{See also http://sites.northwestern.edu/blast/information-for-observers/ for the technical description of the BLAST-TNG experiment (G. Novak, private communication)}]{Fissel13} and the Planck satellite \citep{Pelkonen07}, respectively, and plot it in Fig.~\ref{fig:noise}. For both instruments, $\sigma_{I}$ shows a rough $\lambda^{-2}$ dependence. Hence, we use a noise level of
\begin{equation}
 \sigma_I = 20 \,\rmn{GJy \, sr^{-1}} \times \left( \frac{\lambda}{1~\mu\rmn{m}} \right)^{-2} \, ,
 \label{eq:noise}
\end{equation}
which we consider as a sufficient approximation for our purpose. In a second step, we convolve the noisy data with a Gaussian beam with a width of 3 pixels ($\equiv$ 0.36 pc or 75'' at 1 kpc distance). As shown by \citet{Pelkonen07}, this increases the signal-to-noise ratio and thus improves the quality of the observed polarisation structure. We note that convolving the noisy data with an even wider beam does not significantly improve the results compared to 3 pixel, but would rather smear out strong column density gradients.

\begin{figure*}
 \includegraphics[width=\linewidth]{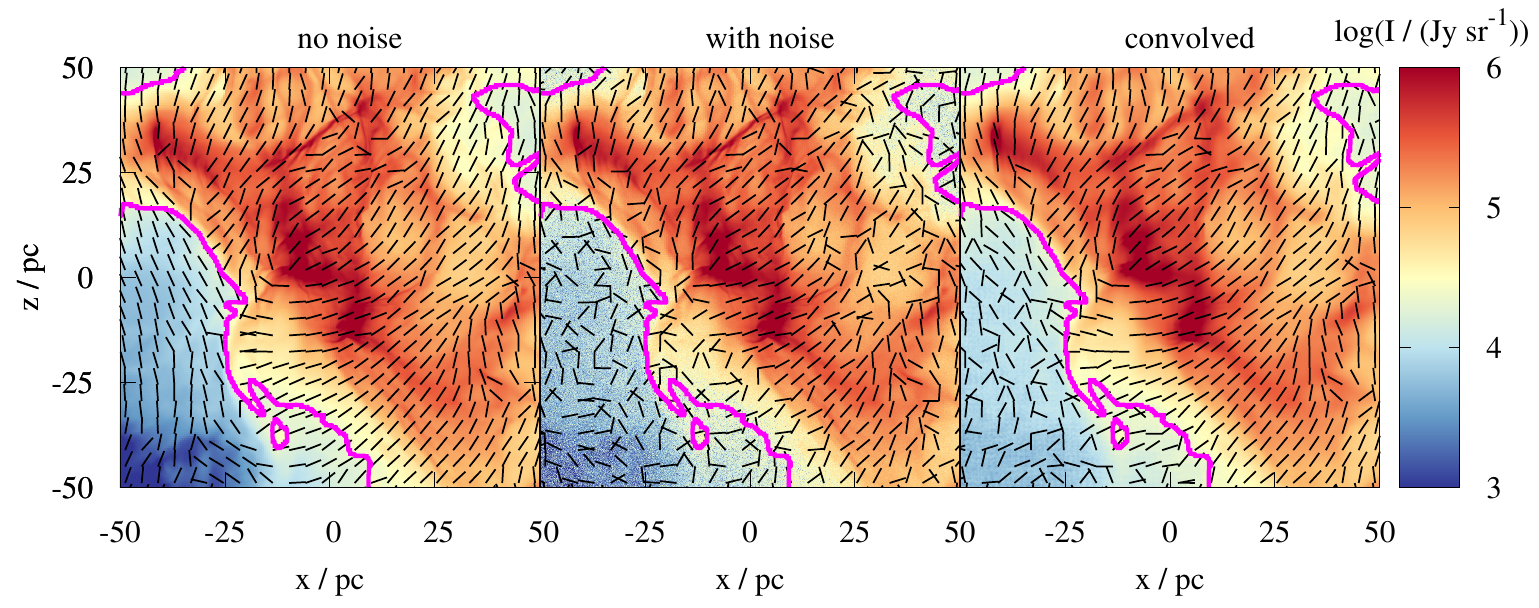}
 \caption{Effect of noise (middle panel) and subsequent convolution (right panel) on the polarisation pattern for MC1 at $t_\rmn{evol}$ = 3 Myr along the $z$-direction. Noise randomizes the polarisation structure compared to the structure without noise (left panel) in regions of low intensity. Convolution with a 3 pixel wide beam (0.36 pc) helps to improve the quality of the polarisation data in regions with intensities above a few times the noise level as indicated by the magenta line showing the 2$\sigma_I$ - contour (Eq.~\ref{eq:noise}). In the high-intensity regions, the polarisation structure is well preserved even for the noisy image.}
 \label{fig:map_noise}
\end{figure*}
In Fig.~\ref{fig:map_noise} we show the intensity and polarisation vectors for MC1 at \mbox{$t_\rmn{evol}$ = 3 Myr} along the $z$-direction at \mbox{$\lambda$ = 1.3 mm.} Adding noise (middle panel) randomizes the polarisation structure, in particular in regions of low intensity \mbox{($I$ $\lesssim$  10$^5$ Jy sr$^{-1}$)}, whereas in the high-intensity regions the polarisation structure is mostly preserved. After convolution (right panel), the polarisation structure is more ordered compared to the unconvolved, noisy image, in particular in regions with $I$ $>$ 2 -- 3 $\times$ \mbox{10$^4$ Jy sr$^{-1}$}. At even lower intensities, however, the vectors remain still unordered with respect to the image without noise.

This value can be related to the noise level added to the map,
\begin{eqnarray}
 \sigma_I(1.3~\rmn{mm}) &=& 20 \, \rmn{GJy \, sr^{-1}} \times \left( \frac{1.3~\rmn{mm}}{1~\mu\rmn{m}} \right)^{-2}   \nonumber \\
  &=& 1.2 \times 10^4 \, \rmn{Jy~sr^{-1}} \, . \nonumber \\
\end{eqnarray}
This is also indicated by the magenta line in Fig.~\ref{fig:map_noise} showing the contour at $2 \sigma_I(1.3~\rmn{mm})$. The contour outlines quite well the region where the convolution significantly improves the quality of the polarisation data.

We find qualitatively similar results also for the remaining clouds and for all other wavelengths except \mbox{$\lambda$ = 70 $\mu$m}, where the observed emission level is lower than the noise level. This is due to the absence of internal heating sources and thus the clouds are IR-dark at 70 $\mu$m (see Appendix~\ref{sec:app_noise} for details). Here, the polarisation structure remains unordered after convolution, which we expect to hold for even shorter wavelengths. For the remaining wavelengths \mbox{$\gtrsim$ 160 $\mu$m}, however, we argue that the polarisation structure can be trusted in regions with \mbox{$I$ $\gtrsim$ 2 $\sigma_I$.} Moreover, also the extent of these regions is very similar (Fig.~\ref{fig:map_noise_lambda}), which is why in the following we stick with the case of \mbox{$\lambda$ = 1.3 mm}.

\begin{figure*}
 \centering
 \includegraphics[width=\linewidth]{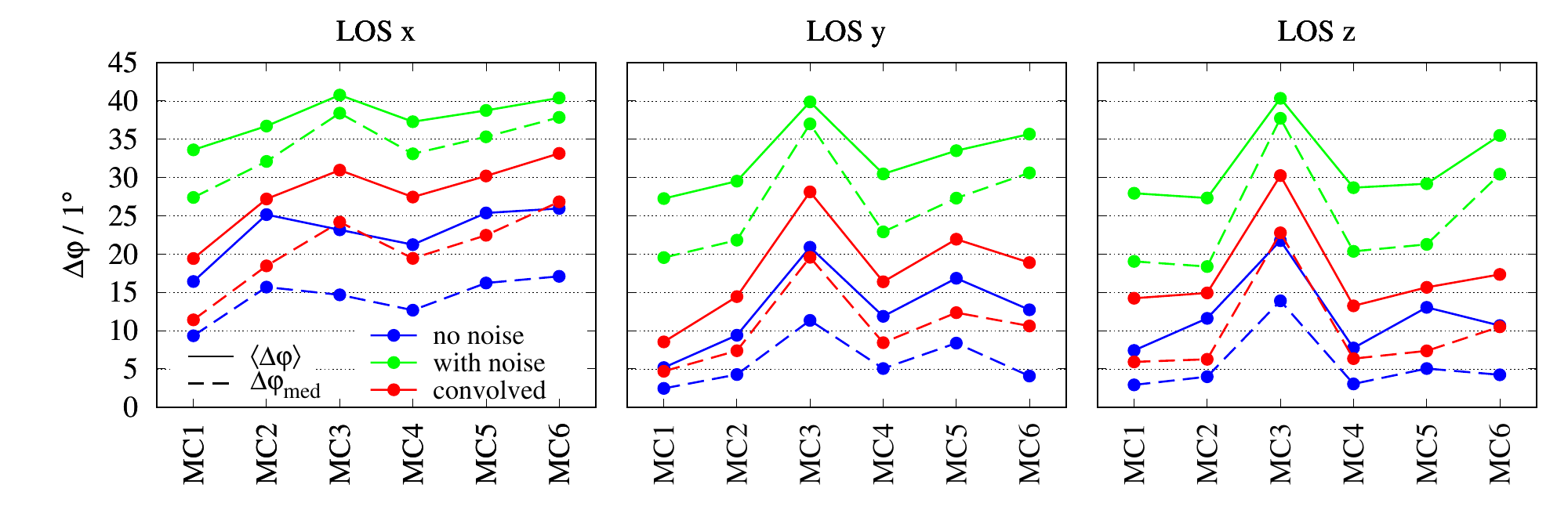}
 \caption{Mean (solid lines) and median deviation (dashed lines) between $\mathbf{B}_\rmn{1.3mm}$ and $\mathbf{B}_\rmn{mw}$ for the six clouds at $t_\rmn{evol}$ = 3 Myr and three different LOS (from left to right). The unconvolved, noisy maps (green lines) typically show 15 -- 20$^\circ$ higher deviations compared to the maps without noise (blue lines), the convolved maps (red lines) only 5 -- 10$^\circ$ higher deviations.}
 \label{fig:dphi_MCs}
\end{figure*}
Next, we calculate the mean and median differences $\left\langle \Delta \varphi \right\rangle$ and $\Delta \varphi_\rmn{med}$ between the observed field and the LOS-averaged magnetic field (see Section~\ref{sec:accuracy}) for all six MCs. We focus on the mass-weighted average $\mathbf{B}_\rmn{mw}$ (Eq.~\ref{eq:Bmw}), as it fits best with the observed field structure (Figs.~\ref{fig:dphi_map} and~\ref{fig:dphi_mean}), and on the snapshot at $t_\rmn{evol}$ = 3 Myr, as the changes over time are rather moderate (Fig.~\ref{fig:dphi_mean}). Overall,  we find qualitatively similar results for all six clouds (see Fig.~\ref{fig:dphi_MCs}): $\Delta \varphi_\rmn{med}$ is smaller than $\left\langle \Delta \varphi \right\rangle$ and the typical deviations are a factor of a few larger along the $x$-direction (left panel) than along the $y$- and $z$-direction (middle and right panel). The typical variations of  $\Delta \varphi_\rmn{med}$ and $\left\langle \Delta \varphi \right\rangle$ among the MCs are of the order of 10$^\circ$.

Adding noise (green lines) typically increases $\Delta \varphi_\rmn{med}$ and $\left\langle \Delta \varphi \right\rangle$ by 15 -- 20$^\circ$ compared to the case without noise (blue lines). For the convolved images (red lines), the deviations are only \mbox{$\sim$ 5 -- 10$^\circ$} above those of the images without noise. Overall, the typical deviations for the convolved images range from \mbox{$\sim$ 5$^\circ$} to 30$^\circ$.

We emphasise that we find similar values for all \mbox{$\lambda$ $\geq$ 161 $\mu$m}. As discussed before, only for $\lambda$ = 70 $\mu$m the field structure remains completely unordered even after the convolution. We note that the relatively high upper limit of $\sim$ 30$^\circ$ for the range of typical variations can be attributed to large regions with low intensities. For this reason, we next discuss criteria to improve the accuracy by identifying regions with a typical deviations of $\lesssim$ 10$^\circ$.

\section{Where can we trust the observed magnetic field structure?}
\label{sec:correlation}

As shown in Fig.~\ref{fig:map_noise}, noise randomizes the observable polarisation structure in regions with $I$ $\lesssim$ 2$\sigma$, which cannot be corrected by convolving the image. Despite the wavelength-dependence of $\sigma_I$, these regions seem to be comparable in their extent (with the only exception of \mbox{$\lambda$ = 70 $\mu$m}, see Fig.~\ref{fig:map_noise_lambda}). For this reason, we try to identify regions with high accuracy and to develop wavelength-independent criteria, which are based on quantities like e.g. the column density or polarisation degree, to assess the reliability of the field structure obtained in actual observations. We discuss this on the basis of our fiducial run MC1 at \mbox{$t_\rmn{evol}$ = 3 Myr} and \mbox{$\lambda$ = 1.3 mm}. The results for the remaining runs and wavelengths, however, are qualitatively and quantitatively similar.

\subsection{Column density}

\begin{figure*}
 \centering
 \includegraphics[width=0.9\textwidth]{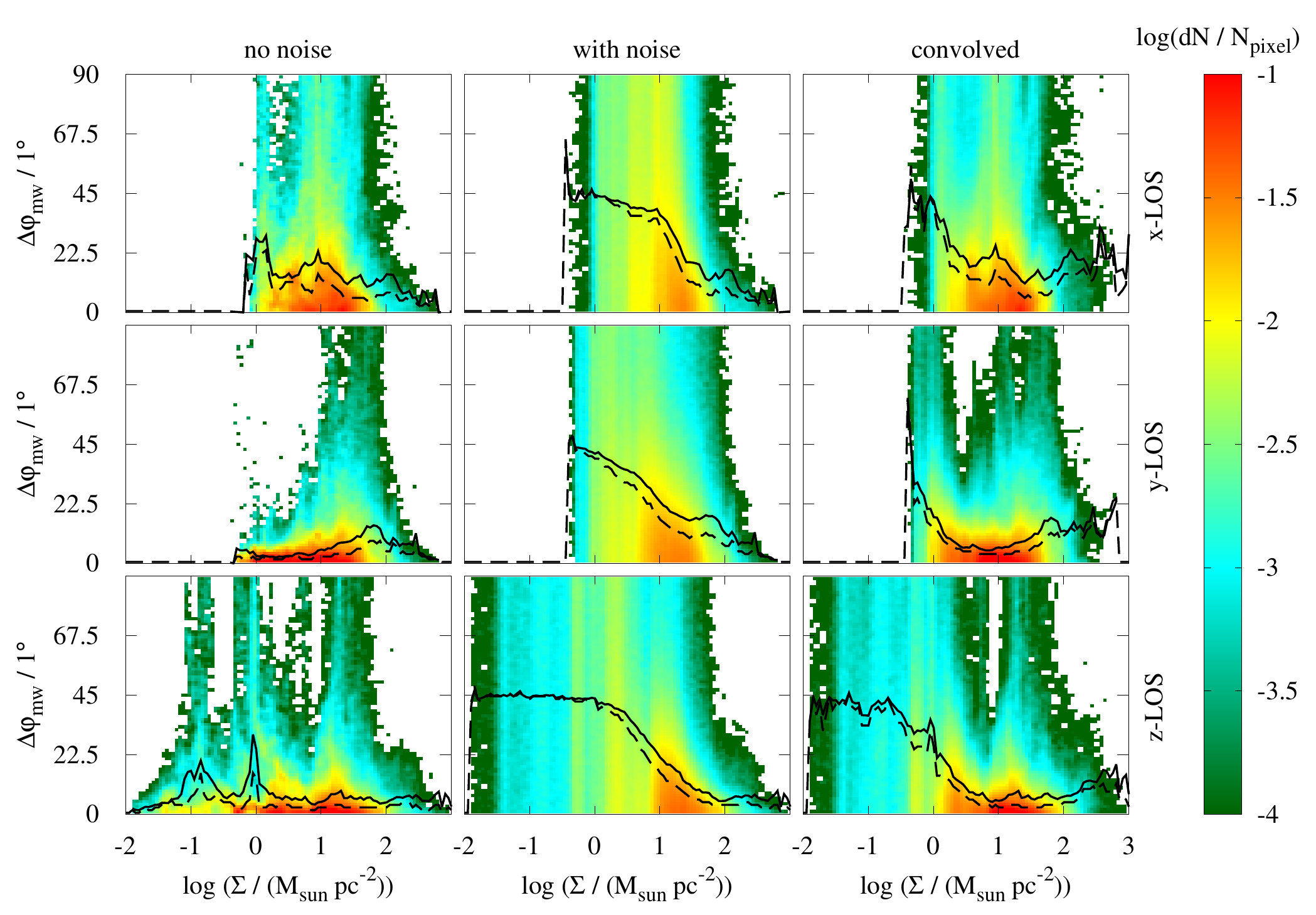}
 \caption{Histogram of the column density and the difference $\Delta \varphi_\rmn{mw}$ between the observed magnetic field orientation at $\lambda$ = 1.3 mm and that of the mass-weighted LOS-averaged magnetic field for MC1 at $t_\rmn{evol}$ = 3 Myr along three different LOS (top to bottom) for the images without and with noise, and the convolved image (from left to right). The solid and dashed black lines show the mean and median of $\Delta \varphi_\rmn{mw}$, respectively. For the convolved image, there is a good match between the observed and the mass-weighted magnetic field at column densities $\Sigma$ $\gtrsim$ 1 M$_{\sun}$ pc$^{-2}$.}
 \label{fig:dphi_cd}
\end{figure*}

In Fig.~\ref{fig:dphi_cd} we show the area-weighted distribution of all pixels in the $\Sigma$ - $\Delta \varphi_\rmn{mw}$ phase space, for the images with (middle) and without noise (left), as well as the convolved, noisy image (right). The column density $\Sigma$ is taken directly from the simulation data. Similar to the Figs.~\ref{fig:dphi_mean} and~\ref{fig:dphi_MCs}, also for the individual $\Sigma$ bins the median of $\Delta \varphi_\rmn{mw}$ (dashed line) is lower than the mean (solid line)

In the absence of noise (left panels), $\Delta \varphi_\rmn{mw}$ and $\Sigma$ are uncorrelated and the magnetic field structure would be probed equally well with an accuracy of about 10$^\circ$ in both the surroundings as well as the central high-density regions of MCs. However, when adding noise (middle panels), $\Delta \varphi_\rmn{mw}$ increases significantly, in particular in the low-$\Sigma$ regime. At \mbox{$\Sigma$ $\lesssim$ 10 M$_{\sun}$ pc$^{-2}$}, the mean and median exceed 20$^\circ$, reaching up to 45$^\circ$. This latter value corresponds to an observed field $\mathbf{B}_\rmn{1.3mm}$ which is completely random and has no relation to the underlying field structure $\mathbf{B}_\rmn{mw}$ any more. For $\Sigma$~$\gtrsim$ a few \mbox{10 M$_{\sun}$ pc$^{-2}$}, however, noise has little effect on $\Delta \varphi_\rmn{mw}$. 

For the convolved images (right column), $\Delta \varphi_\rmn{mw}$ is decreased in particular at \mbox{$\Sigma$ $\gtrsim$ 1 M$_{\sun}$ pc$^{-2}$} and is comparable to that without noise. Here the actual (LOS-averaged) field structure is probed by the observable field with an accuracy of \mbox{$\lesssim$ 10$^\circ$}. Below 1 M$_{\sun}$ pc$^{-2}$, however, convolution cannot improve the accuracy any further and the deviations from the underlying, actual structure become significant. These regions roughly coincide with those where $I$ $<$ 3 $\sigma_I$ (compare Figs.~\ref{fig:dphi_map} and~\ref{fig:map_noise}).

Hence, under realistic observing conditions, the underlying magnetic field structure in pristine MCs can be well probed (deviations $\lesssim$ 10$^\circ$) at column densities above \mbox{$\sim$ 1 M$_{\sun}$ pc$^{-2}$} or a corresponding visual extinction \mbox{$A_\rmn{V}$ $\simeq$ 0.1}. These threshold values seem to be relatively independent of the wavelength and thus the noise level considered (compare magenta lines in the bottom panels of Fig.~\ref{fig:map_noise_lambda}).

\subsection{Column density gradient}

\begin{figure*}
 \centering
 \includegraphics[width=0.9\textwidth]{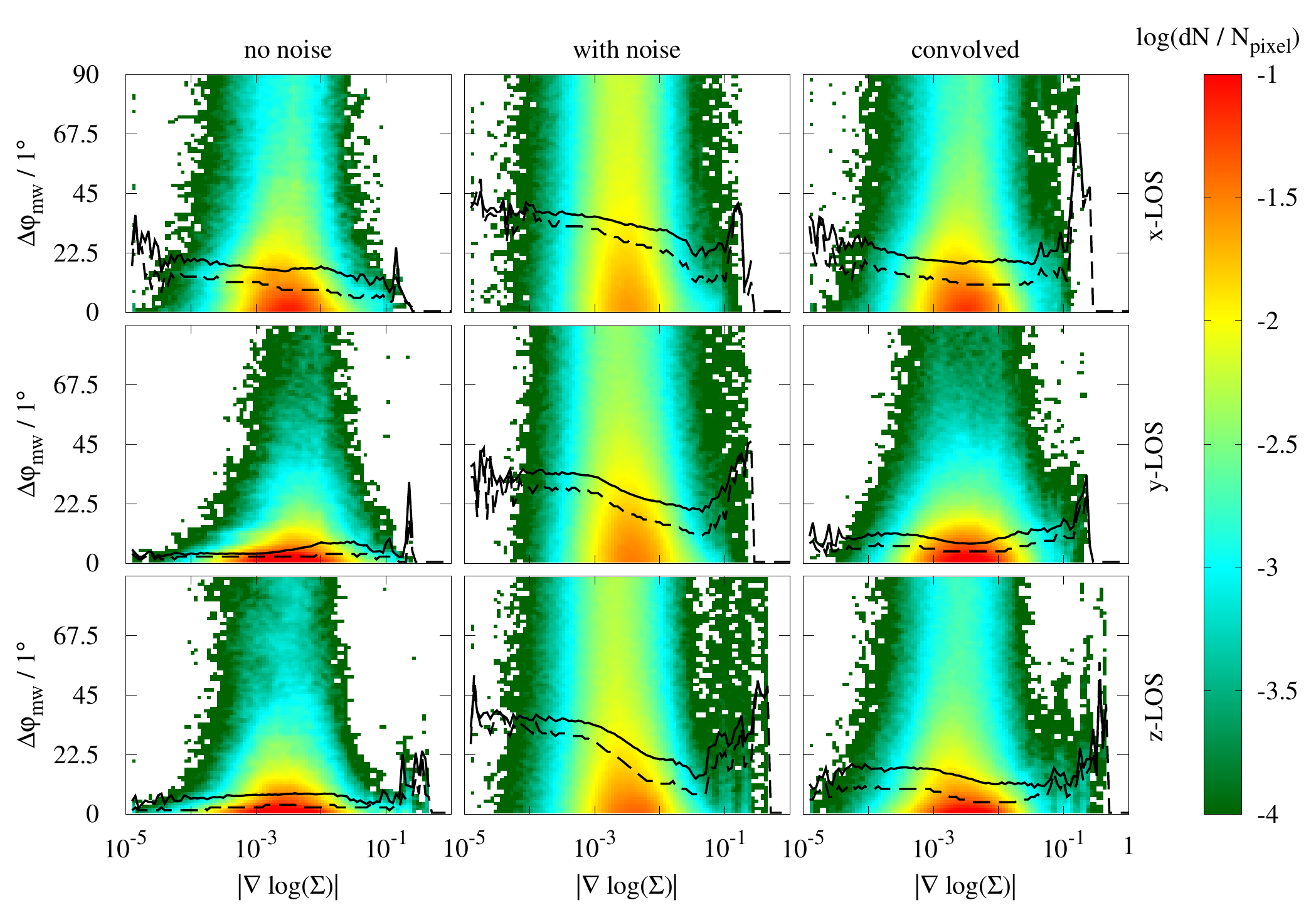}
 \caption{Same as in Fig.~\ref{fig:dphi_cd} but now for the gradient of the logarithmic column density. There is a slight tendency of an increase in $\Delta \varphi_\rmn{mw}$ in regions of strong shocks ($|\nabla \textrm{log}(\Sigma)| > 0.1$).}
 \label{fig:dphi_gradcd}
\end{figure*}

The reliability of the observed field structure might also change in regions with high (column) density variations, e.g. in shocks. Since we are interested in \textit{relative} changes of $\Sigma$, we calculate the gradient of log($\Sigma$), $|\nabla \textrm{log}(\Sigma)|$ using a 7 $\times$ 7 boxcar operator and normalise it to the size of one pixel. Hence, $|\nabla \textrm{log}(\Sigma)|$ = $x$ means that $\Sigma$ changes by a factor of 10$^{x}$ within a distance of 0.12~pc.

We show $|\nabla \textrm{log}(\Sigma)|$ in the bottom middle panel of Fig.~\ref{fig:dphi_map}. As can be seen, the shock-like structure in the lower left corner separating the low-$\Sigma$ environment from the denser parts of the clouds is associated with a significant increase in $\Delta \varphi$ (upper panels). On the other hand, for the region with high $\Delta \varphi$ around ($x$, $y$) $\simeq$ (40 pc, 0 pc) no strong changes in $\Sigma$ are visible.

In order to test this more quantitatively, we show the area-weighted $|\nabla \textrm{log}(\Sigma)|$ - $\Delta \varphi_\rmn{mw}$ phase diagram in Fig.~\ref{fig:dphi_gradcd}. Adding noise and convolving the image shifts $\Delta \varphi_\rmn{mw}$ to overall higher values compared to the case without noise, the functional shapes of $\left\langle \Delta \varphi_\rmn{mw} \right\rangle$ and $\Delta \varphi_\rmn{med,mw}$, however, remain relatively similar. For \mbox{$|\nabla \textrm{log}(\Sigma)| < $ 0.1}, $|\nabla \textrm{log}(\Sigma)|$ and $\Delta \varphi_\rmn{mw}$ are only very weakly anti-correlated and the magnetic field can be probed with an accuracy of $\lesssim$ 10$^\circ$.

For $|\nabla \textrm{log}(\Sigma)|$ $>$ 0.1, however,  $\Delta \varphi_\rmn{mw}$ increases rapidly. Given a pixel size of 0.12 pc, this indicates that the field structure cannot be probed accurately in regions where $\Sigma$ changes by a factor of more than 10$^{0.1}$ $\sim$ 3 within about 0.1~pc. However, as the statistics in this regime is quite low and as the increase is less clear in the case without noise, this findings could partly caused by the added noise. We suggest that higher resolution is required to probe the dependence of the observed magnetic field structure in regions of shocks, as otherwise the field structure might become washed out, an effect which could potentially happen in observations.

\subsection{Polarisation degree}

\begin{figure*}
 \centering
 \includegraphics[width=0.9\textwidth]{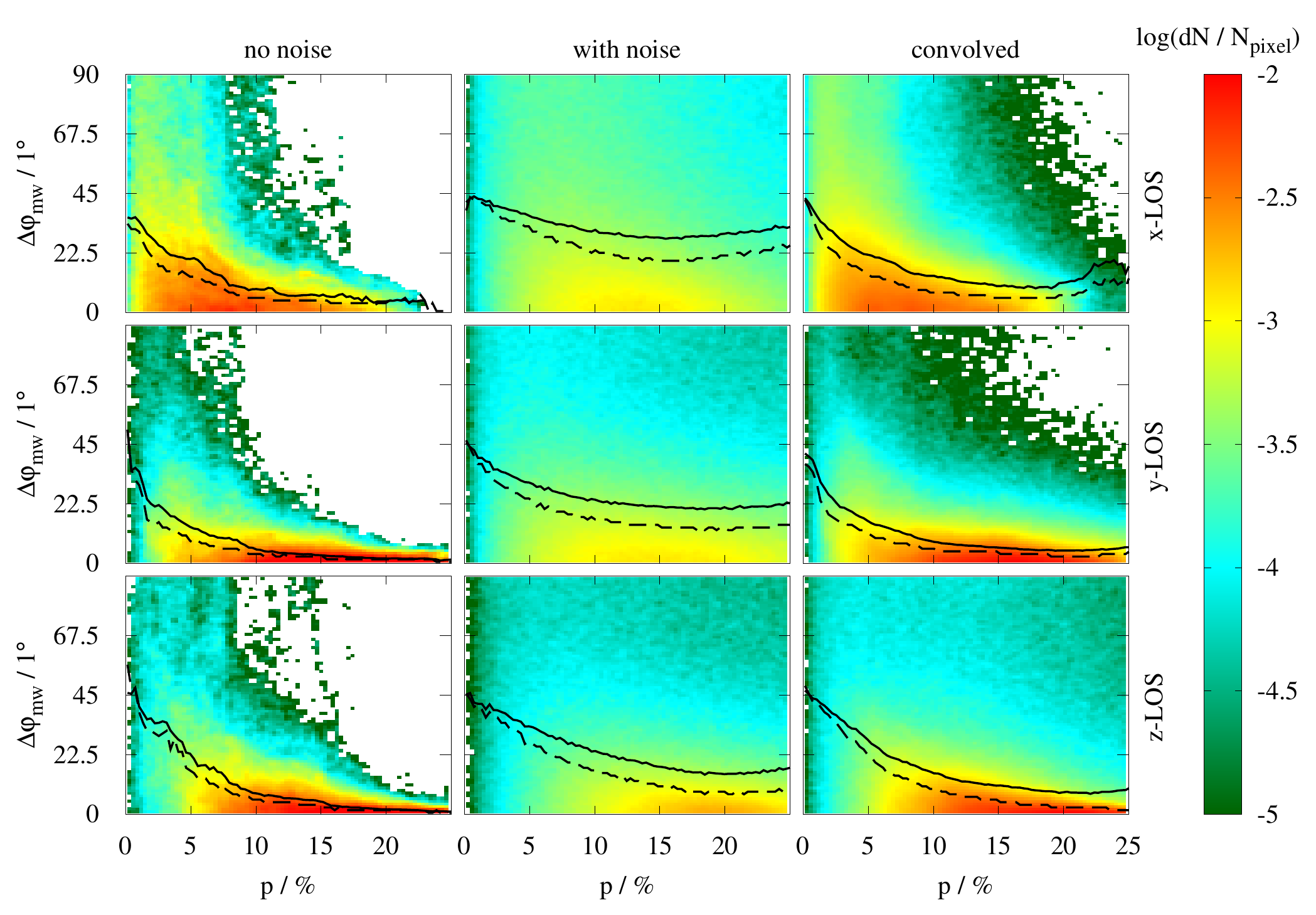}
 \caption{Same as in Fig.~\ref{fig:dphi_cd} but now for the polarisation degree. There is a clear anti-correlation between $\Delta \varphi_\rmn{mw}$ and $p$. In regions with $p$ $\lesssim$ 2\%, the underlying LOS-averaged magnetic field structure cannot be probed reliably.}
 \label{fig:dphi_pol}
\end{figure*}
Comparing the $\Delta \varphi$-maps in the upper row of Fig.~\ref{fig:dphi_map} with the $p$-map in the lower right panel reveals an anti-correlation between $\Delta \varphi$ and $p$: the lower the value of $p$, the less accurate can the LOS-averaged field structure be probed by the observed field structure. This becomes even more clear in the $p$ - $\Delta \varphi_\rmn{mw}$ phase diagrams shown in Fig.~\ref{fig:dphi_pol}. Independently of the presence or absence of noise, $\Delta \varphi_\rmn{mw}$ decreases with increasing $p$. For the convolved image and the image without noise, $\Delta \varphi_\rmn{mw}$ is on average $\sim$ 20$^\circ$ at $p$ = 3 -- 4\% (black lines) and reaches values up to 40$^\circ$ -- 45$^\circ$ at even lower $p$, which corresponds to nearly uncorrelated inferred and LOS-averaged magnetic field orientations. For this reason, we tentatively suggest that $p$ $\simeq$ 2\% is a critical limit, above which one can reliably determine the magnetic field structure from observations.

We note that an even higher threshold for $p$ would be critical since for e.g. \mbox{$p$ = 5\%}, in some cases we would have to dismiss up to 80\% of the pixels, thus significantly decreasing the statistics. We also emphasise that the overall increase in $p$ in the presence of noise is expected: as $p$ is a strictly positive quantity, any noise in $Q$ and $U$ tends to increase the measured value of $p$ \citep[e.g.][]{Wardle74,Fissel16}.

\section{Depolarisation in molecular clouds}
\label{sec:depolarisation}

In recent observations the polarisation degree is often found to decrease towards the dense regions of MCs \citep[e.g.][]{Wolf03,Attard09,Tang09,Bertrang14,Brauer16,Fissel16,Santos17,Galametz18,Koch18}. The cause of this is still under debate \citep[see e.g. the review by][]{Li14}. On the one hand, \citet{Padoan01} suggest that dust grains are not aligned for $A_\rmn{V}$ $>$ 3. On the other hand, strong variations of the magnetic field along the LOS combined with resolution effects can contribute to the depolarisation. The latter was suggested by re-observing regions with low polarisation degrees \citep[found in low-resolution observations by e.g.][]{Attard09,Tang09} with higher resolution \citep[e.g.][]{Girart06,Koch18}. Since we here use a self-consistent model for dust alignment, we can disentangle the relative impact of both effects.

\subsection{Misaligned dust grains?}

\begin{figure}
\includegraphics[width=\linewidth]{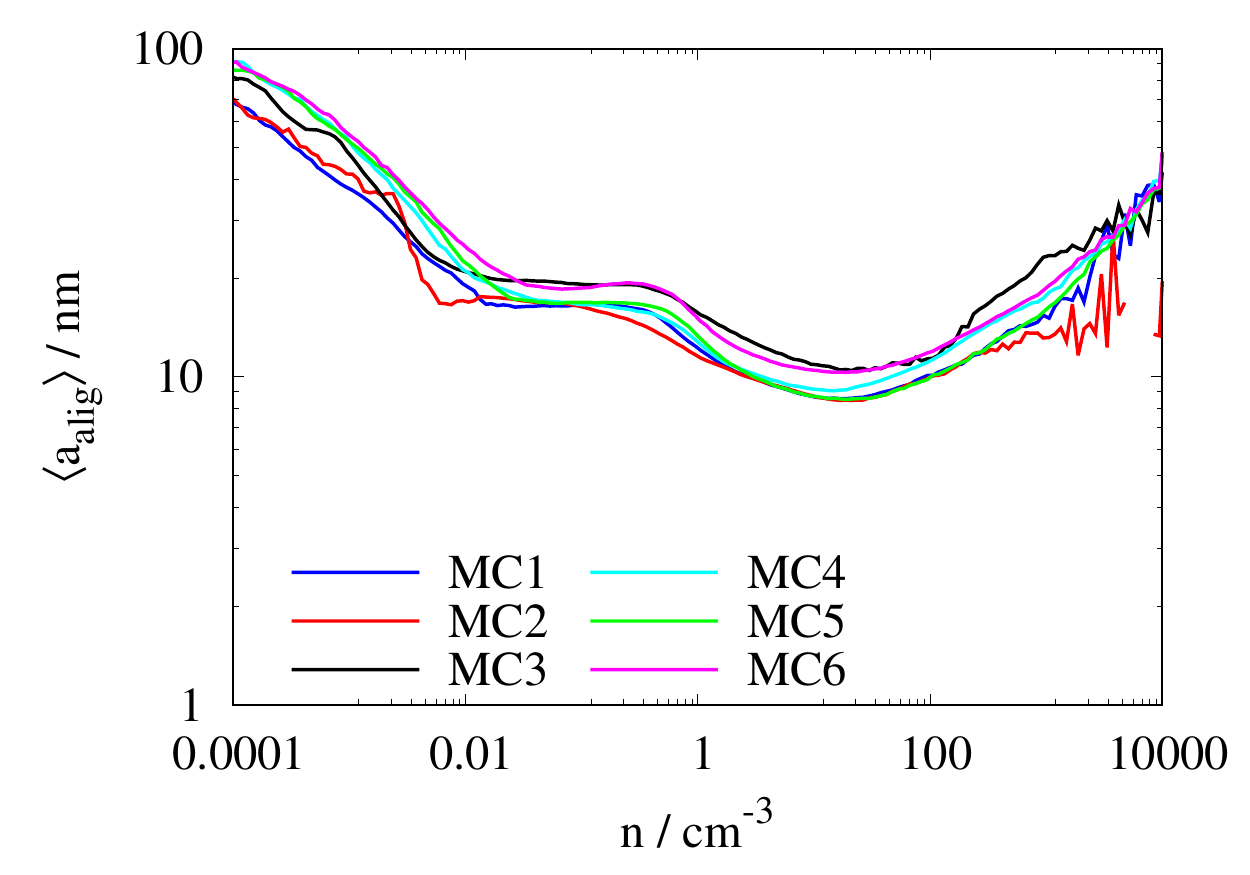}
\caption{Mean value of $a_\rmn{alig}$ (Eq.~\ref{eq:omega}) giving the threshold size above which dust grains are aligned with the magnetic field for all six clouds at $t_\rmn{evol}$ = 3 Myr as a function of density. The dust grains are best aligned in the regime between 1 cm$^{-3}$ and a few 100 cm$^{-3}$, at higher and lower densities the alignment efficiencies decrease.}
\label{fig:aalig}
\end{figure}

As shown in Section~\ref{sec:accuracy}, the observed magnetic field seems to trace best the field structure in the dense regions of MCs. Hence, RAT seems to remain efficient enough to align the dust grains with the magnetic field in these dense regions, despite the shielding of the radiation field by the surrounding gas.

To quantify this further, for each cell we calculate  $a_\rmn{alig}$ (Eq.~\ref{eq:omega}), i.e. the minimum size of the dust grains, which are still aligned with the magnetic field. We do this for all six MCs and plot the mean of $a_\rmn{alig}$, $\left\langle a_\rmn{alig} \right\rangle$, as a function of density in Fig.~\ref{fig:aalig}. At \mbox{$n$ $\lesssim$ 1 cm$^{-3}$}, $\left\langle a_\rmn{alig} \right\rangle$ increases with decreasing $n$ due to the rise in gas temperature (Eq.~\ref{eq:omega}) typically occurring in this density range. The largest fraction of aligned dust grains is found between \mbox{$n$ $\simeq$ 1 cm$^{-3}$} and a few 100 cm$^{-3}$. Here, all dust grains with sizes above \mbox{$\sim$ 10 nm} are aligned with the magnetic field, which we attribute to the significant drop in temperature in this range. Above that density range, $\left\langle a_\rmn{alig} \right\rangle$ starts to increase again to values around 30 -- 40 nm at $n$ = 10$^4$ cm$^{-3}$. This is due to the combined effect of the aforementioned shielding of the radiation field and the simultaneous increase of random gas bombardment ($\propto n$), which leads to a progressive misalignment of small dust grains (Eq.~\ref{eq:omega}). The upper limit $a_l$, up to which grains are aligned (Eq.~\ref{eq:a_l}), remains in our simulations above the maximum grain size of $a_\rmn{max}$ = 2 $\mu$m, and can thus be neglected. Overall we thus see that RAT alignment remains efficient in pristine MCs up to densities of $\sim$ 10$^4$ cm$^{-3}$.

Next, we approximate the fraction of polarised emission, $f_\epsilon$, emitted from each cell under the assumption that the dust grains are in thermal equilibrium, i.e. the absorption cross-section is identical to the emissivity $\epsilon$ of each grain. In the Rayleigh limit, i.e. when the grain size $a$ is smaller than the wavelength, the absorption cross-section, and thus $\epsilon$, is proportional to the \textit{volume} of the grain \citep[e.g.][Chapter~5]{Bohren83}, and thus larger dust grains contribute significantly more to the emission. With $a_\rmn{max}$ = 2 $\mu$m and typical values of \mbox{$\lambda$ $\gtrsim$ 100 $\mu$m}, the  Rayleigh limit provides a reasonable approximation. Hence, $f_\epsilon$ can be determined as
\begin{equation}
 f_\epsilon = \frac{\int_{a_\rmn{alig}}^{a_\rmn{max}} a^{-3.5} \times a^3 \rmn{d}a}{\int_{a_\rmn{min}}^{a_\rmn{max}} a^{-3.5} \times a^3 \rmn{d}a}  \propto \sqrt{a_\rmn{max}} - \sqrt{a_\rmn{alig}} \, ,
\end{equation}
where the nominator serves to normalise $f_\epsilon$. As \mbox{$a_\rmn{alig}$ $<<$ $a_\rmn{max}$ = 2 $\mu$m} (Fig.~\ref{fig:aalig}), $f_\epsilon$ depends only weakly on the actual value of $a_\rmn{alig}$.

Indeed, for our six MCs the values of $f_\epsilon$ are around 90 -- 100\% for the entire density range up to $\sim$ 10$^4$ cm$^{-3}$. Hence, based on this simple estimate we expect also the densest cells to significantly contribute to the polarised emission. Furthermore, as we have shown in \citet{Seifried17} (see their Fig.~12), regions at $n$ $\gtrsim$ 1000 cm$^{-3}$ typically have visual extinctions \mbox{$A_\rmn{V}$ $\gtrsim$ 3}. Our results thus questions the suggestion by \citet{Padoan01} that grain misalignment at \mbox{$A_\rmn{V}$ $\gtrsim$ 3} causes the observed depolarisation. They are, however, still compatible with the observations suggesting alignment to cease at \mbox{$A_V$ $\gtrsim$ 20} \citep{Alves14,Jones15}.

To summarise, dust grains remain well aligned with the magnetic field even in high density regions ($n$ $\gtrsim$ 10$^3$ cm$^{-3}$, $A_\rmn{V}$ $>$ 1), which are dominating the observed polarisation causing the observed field to trace the mass-weighted magnetic field (Section~\ref{sec:accuracy}). However, at densities higher than $\sim$ 10$^4$ cm$^{-3}$ not covered in our simulations, the alignment efficiencies most likely decrease, which can contribute to the frequently observed dust polarisation holes in dense protostellar cores.

\subsection{Magnetic field variations along the LOS?}
\label{sec:LOSvariations}

\begin{figure}
 \centering
 \includegraphics[width=\linewidth]{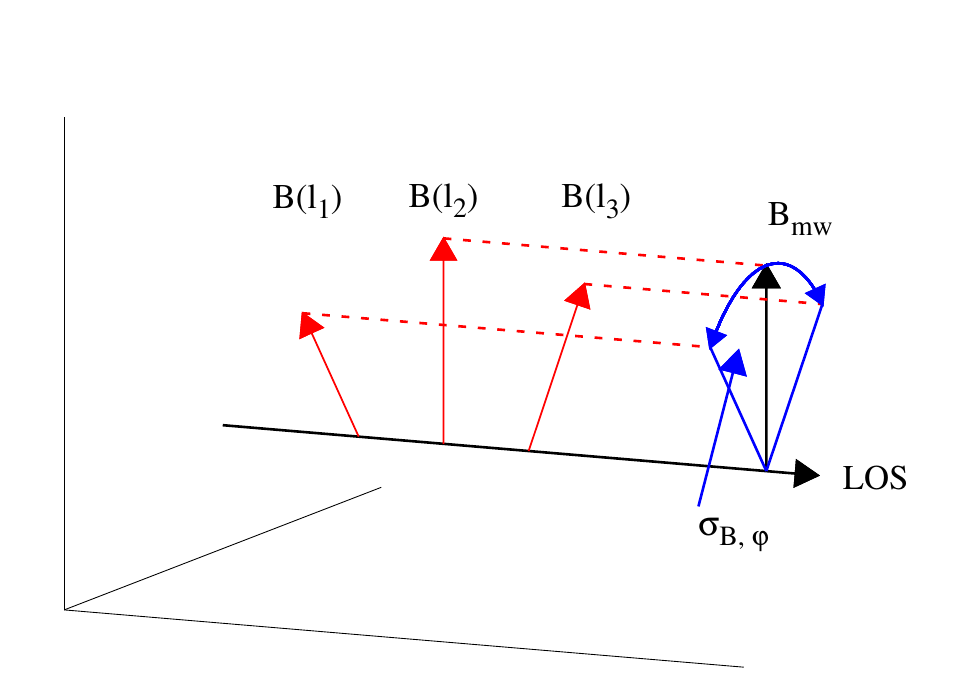}
 \caption{Sketch of $\sigma_{B, \varphi}$ (Equation~\ref{eq:sigma}, blue arrows), which is a measure for the typical variation of the magnetic field along the LOS (red arrows) around the LOS-averaged field (black arrow).}
 \label{fig:sketch}
\end{figure}
Next, we investigate tangled magnetic field lines as a reason of low polarisation degrees. For this purpose, we calculate the variation of the magnetic field orientation along a given LOS around the direction of the LOS-averaged field $\mathbf{B}_\rmn{mw}$ (Eq.~\ref{eq:Bmw}, see Fig.~\ref{fig:sketch} for a graphical representation), i.e.
\begin{equation}
 \sigma_{B, \varphi} = \frac{\int_\rmn{LOS} \rho \, \measuredangle (\mathbf{B}(l), \, \mathbf{B}_\rmn{mw}) \rmn{d}l}{\int_\rmn{LOS} \, \rho \, \rmn{d}l} \, .
 \label{eq:sigma}
\end{equation}

\begin{figure}
 \includegraphics[width=\linewidth]{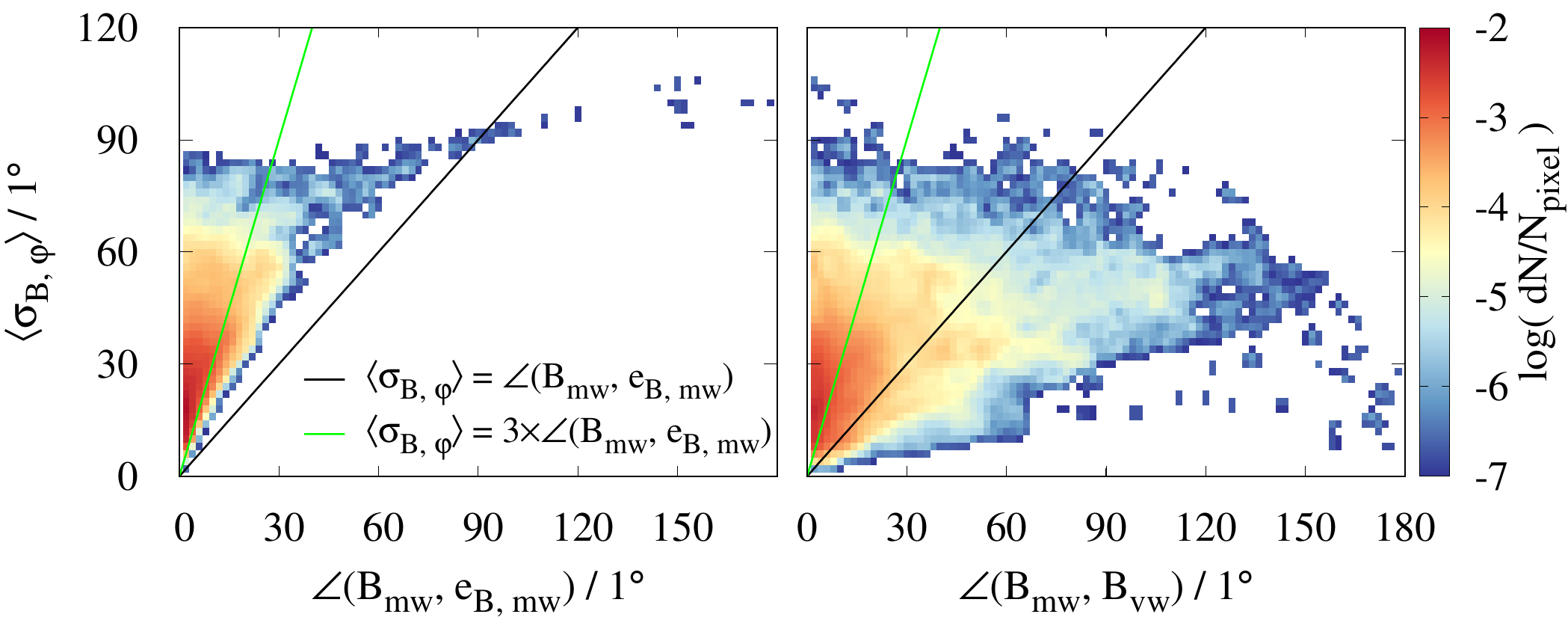}
 \caption{Phase diagram of the LOS variation $\sigma_{B, \varphi}$ versus $\measuredangle (\mathbf{B}_\rmn{mw},\mathbf{e}_\rmn{B,mw})$ (left) and $\measuredangle (\mathbf{B}_\rmn{mw},\mathbf{B}_\rmn{vw})$ (right) for MC1 at $t_\rmn{evol}$ = 3 Myr along the $z$-direction. To guide the readers eye we show the lines where $\sigma_{B, \varphi}$ is 1 times (black) and 3 times (green) the angle between the definitions. For the vast majority of points, $\sigma_{B, \varphi}$ is a factor of a few larger than the differences obtained by using different definitions.}
 \label{fig:angle_diff}
\end{figure}

We use the mass-weighted magnetic field vector as a reference as it best fits the orientation of the polarisation. However, as discussed in Section~\ref{sec:accuracy}, there are various ways to define the mean magnetic field. In order to investigate how significant the LOS variations are with respect to the differences between the various definitions, we plot $\sigma_{B, \varphi}$ against the angle between $\mathbf{B}_\rmn{mw}$ and $\mathbf{e}_\rmn{B,mw}$, $\measuredangle (\mathbf{B}_\rmn{mw},\mathbf{e}_\rmn{B,mw})$, as well as $\mathbf{B}_\rmn{vw}$, $\measuredangle (\mathbf{B}_\rmn{mw},\mathbf{B}_\rmn{vw})$, in Fig.~\ref{fig:angle_diff} for MC1 at $t_\rmn{evol}$ = 3 Myr along the $z$-direction. We emphasise that we get qualitative and quantitative very similar results for all remaining clouds and directions.

As can be seen, $\measuredangle (\mathbf{B}_\rmn{mw},\mathbf{B}_\rmn{vw})$ (right panel) shows a larger scatter and tends to somewhat higher values than $\measuredangle (\mathbf{B}_\rmn{mw},\mathbf{e}_\rmn{B,mw})$ (left panel). Looking at the different clouds and directions, we find that typically the mean of $\measuredangle (\mathbf{B}_\rmn{mw},\mathbf{e}_\rmn{B,mw})$ is around 10$^\circ$ $\pm$ 5$^\circ$, whereas the mean of $\measuredangle (\mathbf{B}_\rmn{mw},\mathbf{B}_\rmn{vw})$ is around 15$^\circ$ $\pm$ 10$^\circ$. Hence, despite some scatter, the LOS averages agree reasonably well for the different definitions for the majority of the pixels. Moreover, $\sigma_{B, \varphi}$ is a factor of a few larger than $\measuredangle (\mathbf{B}_\rmn{mw},\mathbf{e}_\rmn{B,mw})$ and $\measuredangle (\mathbf{B}_\rmn{mw},\mathbf{B}_\rmn{vw})$ as indicated by the black and green lines. With other words: the variation $\sigma_{B, \varphi}$ along the LOS is typically larger than the uncertainty introduced by different LOS averaging procedures and can thus be considered as a significant and meaningful quantity to assess the order of the magnetic field along the LOS.

\begin{figure}
 \centering
 \includegraphics[width=\linewidth]{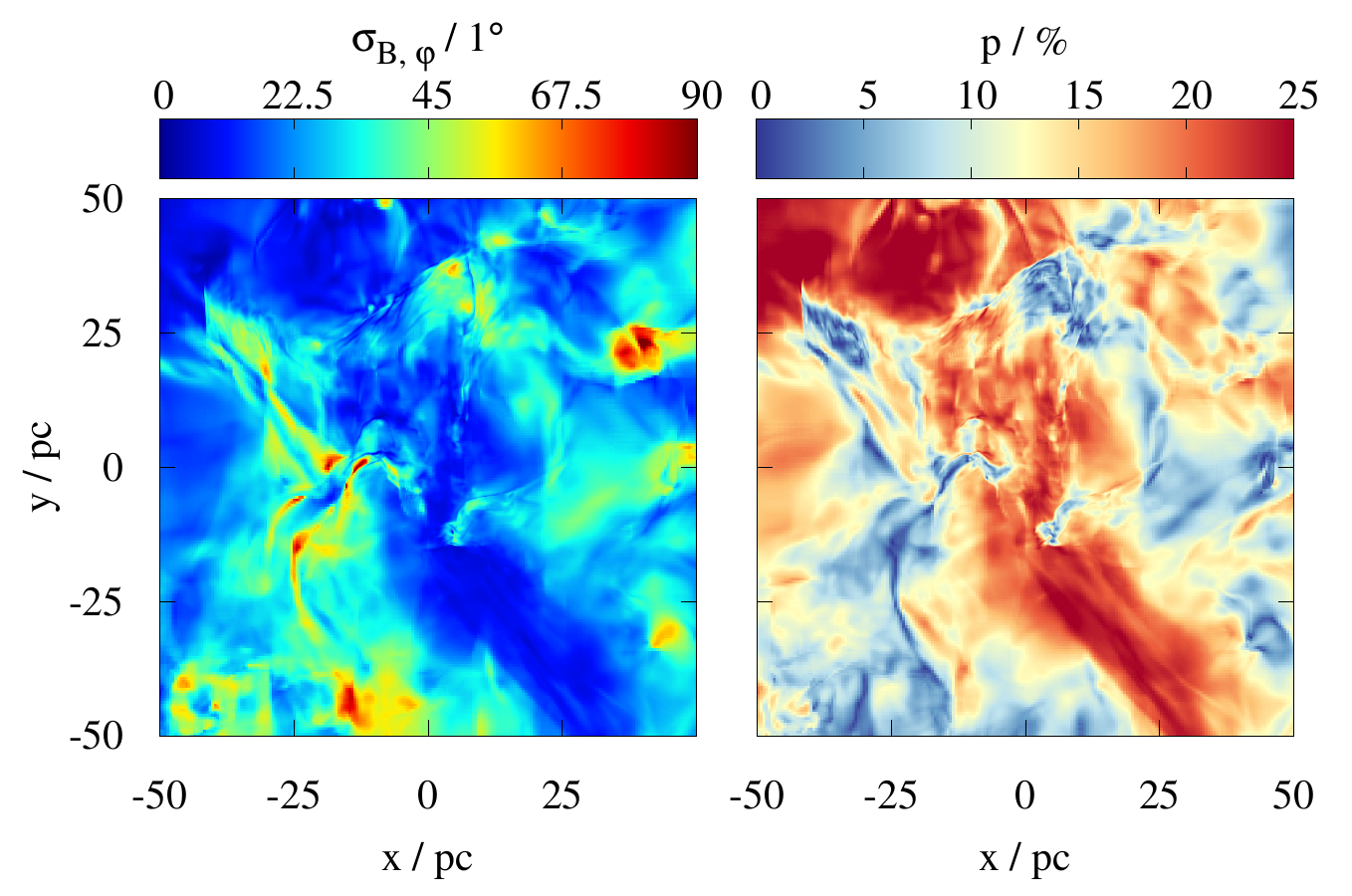}
 \caption{Maps of $\sigma_{B, \varphi}$ (left panel, see Eq.~\ref{eq:sigma}) and $p$ (right panel) for MC1 at $t_\rmn{evol}$ = 3 Myr along the $z$-direction. There exists a correlation between $p$ and $\sigma_{B, \varphi}$.}
 \label{fig:map_sigma}
\end{figure}

We plot the resulting $p$ and $\sigma_{B, \varphi}$ maps for MC1 at \mbox{$t_\rmn{evol}$ = 3 Myr} along the $z$-direction in Fig.~\ref{fig:map_sigma}, where we have taken $p$ for the case without noise. There is a clear anti-correlation between $p$ and $\sigma_{B, \varphi}$ visible: The more the orientation of the magnetic field varies along a given LOS, the lower is the resulting polarisation degree. This is due to the fact that with a varying field direction also the orientation of the dust grains changes along the LOS, thus lowering the net polarisation.

\begin{figure}
 \centering
 \includegraphics[width=\linewidth]{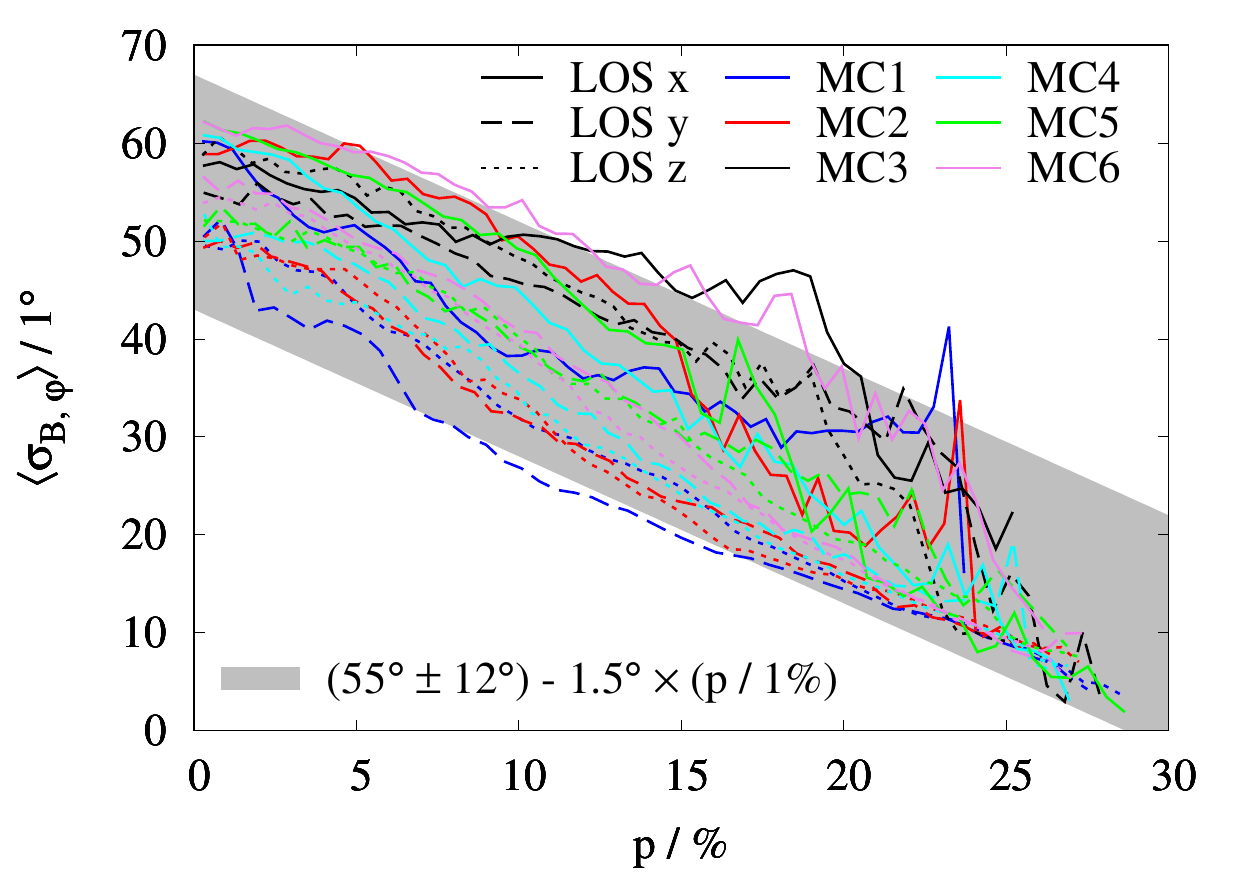}
 \caption{Typical variation $\left\langle \sigma_{B, \varphi} \right\rangle$ of the magnetic field orientation along a LOS at a given polarisation degree $p$ for different clouds (colour coded) and different LOS (solid, dashed, and dotted). There is a roughly linear correlation indicated by the grey-shaded area.}
 \label{fig:sigma_p}
\end{figure}

Next, for a given $p$ we calculate the mean variation $\left\langle \sigma_{B, \varphi} \right\rangle$ of the magnetic field along the LOS averaged over the entire map for all six MCs and all three direction. For most of the cases, there is a clear anti-correlation between $p$ and $\left\langle \sigma_{B, \varphi} \right\rangle$ (Fig.~\ref{fig:sigma_p}), which seems to follow a rough global trend. This would allow us to infer the typical variation of the magnetic field along the considered LOS from the observed polarisation degree. For our simulated MCs we find a roughly linear relation (grey shaded area) of 
\begin{equation}
 \left\langle \sigma_{B, \varphi} \right\rangle \simeq (55^\circ \pm 12^\circ) - 1.5^\circ \times \frac{p}{1\%} \, ,
\end{equation}
which might be a useful approximation in future observations. This relation implies that for \mbox{$p$ $\lesssim$ 2\%} the observed field does not reliably reproduce the underlying 3D structure of the field due to its strong variations along the LOS. This is in agreement with our findings that in this regime $\Delta \varphi_\rmn{mw}$ can be significant (see Fig.~\ref{fig:dphi_pol}).

To summarize, our results indicate that on MC scales the variation of the magnetic field along the LOS significantly contributes to the depolarisation instead of a decrease in alignment efficiencies. Performing a test RT calculation assuming that all dust grains are perfectly aligned, i.e. artificially excluding grain misalignment as a source of depolarisation, we also find an anti-correlation between $p$ and $\sigma_{B, \varphi}$, which further strengthens our conclusion.

We note that also a magnetic magnetic field orientation which is preferentially parallel to the LOS can reduce the overall polarisation degree. This is visible in Fig.~\ref{fig:dphi_pol}, in particular for the case without noise (left column), where $p$ appears to be slightly lower along the $x$-direction, i.e. parallel to the orientation of the original field.

\section{Discussion}
\label{sec:discussion}

\subsection{Comparison with a simplified method}

In this work we used MHD simulations of MC formation, which include the shielding of UV radiation by gas and dust, and applied fully self-consistent dust polarisation RT calculations with a self-consistent description of dust alignment efficiencies \citep{Lazarian07review,Andersson15}. For this reason, we are able to test how well simpler schemes match the observable polarisation structure.

In the following we test a simplified method to predict the observable polarisation structure from simulations which is based on works of \citet{Lee85}, \citet{Wardle90}, and \citet{Fiege00}, and has been applied in a number of publications \citep{Heitsch01,Padoan01,Kataoka12,Soler13,PlanckXX,King18,Vaisala18}. It is also the basis of the DustPol module \citep{Padovani12} contained in the RT code ARTIST \citep{Brinch10,Jorgensen14}. Although this method is used widely in literature, a proof-of-concept is still lacking. In the following we briefly describe the method, for more details we refer to the aforementioned publications.
 
For the sake of simplicity, we assume that our LOS is along the $z$-direction and the $y$-direction is pointing upwards. With this, the Stokes parameters can be approximated as
\begin{equation}
I = \int S_{\nu}e^{-\tau_{\nu}} \left(1 - p_0\left(\frac{B_x^2 + B_y^2}{B^2} - \frac{2}{3}\right) \right) \rmn{d}l \, ,
\label{eq:Irad}
\end{equation}
\begin{equation}
Q = p_0 \int S_{\nu}e^{-\tau_{\nu}} \left(\frac{B_y^2 - B_x^2}{B^2}\right) \rmn{d}l \, ,
\label{eq:Qrad}
\end{equation}
\begin{equation}
U = p_0 \int S_{\nu}e^{-\tau_{\nu}}\left(\frac{2B_xB_y}{B^2}\right) \rmn{d}l \, ,
\label{eq:Urad}
\end{equation}
where $S_{\nu}$ is the source function and $\tau_{\nu}$ the optical depth. An essential quantity here is the parameter $p_0$ for the intrinsic polarisation fraction, which is a  measure for the alignment efficiency of the dust. It is usually set to a constant value of \mbox{$p_0$ $\simeq$ 0.15 -- 0.2} throughout the simulation domain. Hence, it neglects the impact of varying gas densities and radiation intensities, which would enter via $a_\rmn{alig}$ and $a_\rmn{l}$ (Eqs.~\ref{eq:omega} and~\ref{eq:a_l}), and thus represents a rather crude approximation.

Assuming now optically thin emission ($\tau_{\nu}$ $<<$ 1) and that $S_{\nu}$ is proportional to the radiation intensity of a blackbody with a constant dust temperature throughout the simulation domain, we can rewrite Eqs.~\ref{eq:Irad} -- \ref{eq:Urad} by means of a column density integral
\begin{equation}
\widetilde{I}  = \int \rho \left(1 - p_0\left(\frac{B_x^2 + B_y^2}{B^2} - \frac{2}{3}\right) \right) \rmn{d}l \, ,
\label{eq:Irho}
\end{equation}
\begin{equation}
\widetilde Q = p_0 \int \rho \left(\frac{B_y^2 - B_x^2}{B^2}\right) \rmn{d}l \, ,
\label{eq:Qrho}
\end{equation}
\begin{equation}
\widetilde U = p_0 \int \rho \left(\frac{2B_xB_y}{B^2}\right) \rmn{d}l \, .
\label{eq:Urho}
\end{equation}

\begin{figure*}
 \centering
 \includegraphics[width=\linewidth]{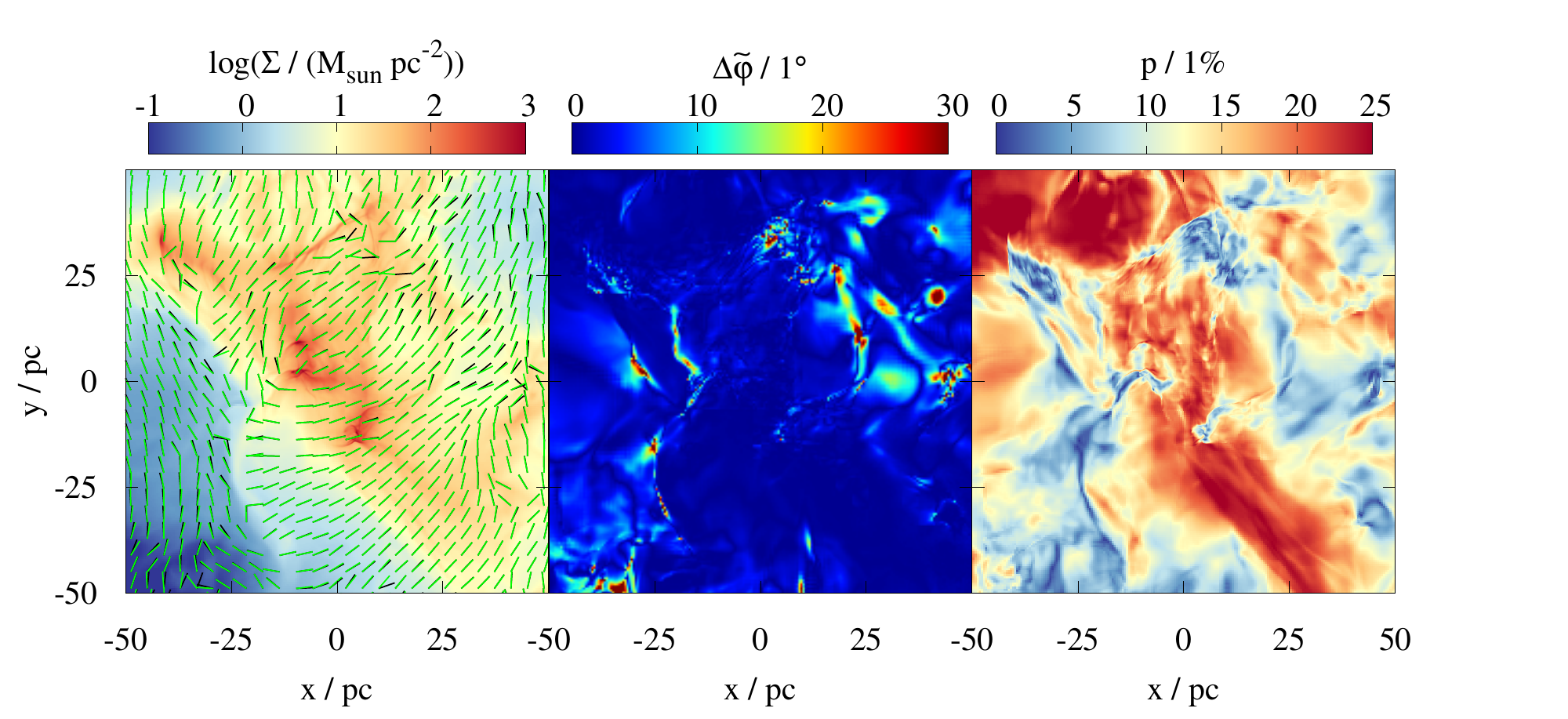}
 \caption{Left: Column density of MC1 at $t_\rmn{evol}$ = 3 Myr along the $z$-direction overplotted with the polarisation vectors calculated with the full RT and self-consistent alignment efficiencies (black) and the simplified method (Eqs.~\ref{eq:Irho} -- \ref{eq:Urho}, green). Middle: Map of the angle residual of both vectors. Overall, the simplified method models the observable structure well with an average residual of a few 1$^\circ$. A larger residual is often found in regions with low polarisation degrees (right panel).}
 \label{fig:dphi_planck}
\end{figure*}
Using $\widetilde Q$ and $\widetilde U$, one can now calculate the polarisation angle $\widetilde{\varphi}$ (Eq.~\ref{eq:p}). Note that since $p_0$ is assumed to be constant, it cancels out for the calculation of $\widetilde{\varphi}$ and does not have to be set here. Next, we compare $\widetilde{\varphi}$ to the polarisation angle obtained via our 1.3 mm RT calculations, $\varphi_\rmn{1.3mm}$, in absence of any noise. For this purpose, we calculate the residual between both angles in each pixel \citep[see Eq.~7 in][]{PlanckXIX} for two different sets of Stokes parameters:
\begin{equation}
 \Delta \widetilde{\varphi} = \widetilde{\varphi} - \varphi_\rmn{1.3mm} \,  \equiv \frac{1}{2}\text{arctan}(Q \widetilde U - U \widetilde Q,\, Q \widetilde Q + U \widetilde U) \, ,
\end{equation}
where $Q$ and $U$ are the Stokes parameters at 1.3 mm obtained from POLARIS. We show the resulting map for MC1 at $t_\rmn{evol}$ = 3 Myr along the $z$-direction in Fig.~\ref{fig:dphi_planck}. The polarisation vectors obtained with both methods (left panel) differ only little resulting in overall small residual values $\Delta \widetilde{\varphi}$ ($\lesssim$ 10$^\circ$, middle panel). Hence, the simplified method seems to represent the observable polarisation structure (in the ideal case without noise) with a good accuracy. There are, however, regions, where the residual is larger. By comparing with the corresponding $p$-map (right panel), we find a slight tendency of increasing $\Delta \widetilde{\varphi}$ with decreasing $p$, i.e. at those positions with significant variations of the field direction along the LOS (see Fig.~\ref{fig:sigma_p}). Repeating this analysis for all six MCs and three LOS, we find that the mean residual is around $\left\langle \Delta \widetilde{\varphi} \right\rangle$ $\simeq$ 1 - 5$^\circ$. Hence, on average Eqs.~\ref{eq:Irho} -- \ref{eq:Urho} seem to provide a good approximation for the actually  observable polarisation structure expect in regions with very low polarisation degrees.

Despite the good match between the simplified and self-consistent model concerning the polarisation \textit{structure}, the simplified model cannot be used to self-consistently calculate the polarisation \textit{degree} since the required alignment efficiencies $p_0$ are set to a constant value independent of the dust and gas properties. Furthermore, a constant dust temperature is assumed and no RT is applied. The latter thus ignores potential optical depth effects, which, as shown in \citet{Brauer16}, can suppress linear polarisation in high-density cloud cores. Moreover, potential wavelength effects are ignored, in particular at $\lambda$ $<$ 70 $\mu$m, where the transition from thermal re-emission to dichroic extinction occurs as e.g. visible for protostellar outflows \citep{Reissl17}. Hence, in this regime the simplified approach is not applicable. Also towards high-density clouds cores, where the alignment efficiency is expected to drop (see Section~\ref{sec:depolarisation}), the assumption of a fixed $p_0$ might strongly limit the applicability of this method.

We note that the DustPol method \citep{Brinch10,Padovani12,Jorgensen14} applies a full RT calculation but still lacks the self-consistent determination of $p_0$. Conversely, \citet{Pelkonen07,Pelkonen09} apply an empirical formula for $p_0$ given by \citet{Cho05}, but only use an approximate way to calculate the Stokes parameters $I$, $Q$, and $U$. Furthermore, none of the afore discussed models can account for circular polarisation, which is calculated automatically with POLARIS. Although we do not discuss circular polarisation in this paper, it might help to disentangle the effects of LOS averaging of magnetic fields \citep{Reissl14}.

\subsection{Caveats}
\label{sec:caveats}

Throughout this work,  we assume a fixed dust-to-gas ratio everywhere in the simulation domain. As shown by \citet{Roman17}, there might be variations by a factor of $\sim$ 3 from the diffuse to the dense ISM. On the other hand, recent theoretical work by \citet{Tricco17} show that in the dense ISM itself (as considered here), the dust-to-gas ratio might not vary by more than a few 10\%, which is why we consider our approach as sufficient.

Furthermore, we keep the dust grain size distribution as well as the ratio between graphite and silicate materials constant (0.375 to 0.625) and do not consider potential effects of ice mantles. Given the wide range of densities and temperatures as well as different histories of the dust grains \citep[e.g.][]{Zhukovska16,Peters17}, we expect the grain properties to change for different cells and LOS. Indeed, recent results of the Planck mission \citep{PlanckXI,Fanciullo15,PlanckXXII,PlanckXXIX} as well as of the BlastPol experiment \citep{Ashton18,Santos17} have shown that the properties of dust grains can vary significantly for different LOS. In particular, dust growth is discussed as an additional reason for depolarisation in MCs \citep[e.g.][]{Bethell07,Pelkonen09,Brauer16} beside LOS variations and dust misalignment (Section~\ref{sec:depolarisation}). Its net effect, however, is unclear as larger grains are better aligned than smaller grains (Eq.~\ref{eq:omega}) but are also contributing less to the polarisation. Performing a first test using $a_\rmn{max}$ = 2~mm showed that in this case the polarisation degree indeed drops whereas the polarisation pattern remains unchanged. Furthermore, taking dust grain aspect ratios smaller than \mbox{$s$ = 0.5} might cause regions with high polarisation degree to become slightly more narrow, the minimal and maximal values of $p$, however, would remain roughly unchanged.

As all these effects can mutually influence each other, detailed RT calculations with varying dust properties would be required. However, since we are here interested in a general proof-of-concept of the applied method, we postpone the effect of varying dust properties to upcoming work.

We also note that more sophisticated methods to correct for noise can be used when calculating the polarisation degree and angle \citep{Wardle74,Witzel11,Montier15}. It is, however, questionable whether these methods would lead to a significant improvement in regions with a low signal-to-noise ratio ($I$ $\lesssim$ 3 $\sigma_I$), where the polarisation pattern is almost random (lower left corner in the panels of Fig.~\ref{fig:map_noise}). Given the improvement achieved by simply convolving the noisy images, makes us confident that this approach is sufficient for our purpose.

As stated before, in this work we only consider pristine MCs, i.e. MCs before the onset of stellar feedback. However, stellar feedback is found to have a significant dynamical impact on clouds \citep[see e.g. the reviews by][]{Krumholz14,Dale15}. It generates H{\sc ii} regions and wind-blown bubbles in the MCs, thereby partly compressing gas as well as the frozen-in magnetic field lines into thin shells, which in turn alters the overall magnetic field morphology. We note that we investigate the general effect of stellar feedback on the dynamics of these clouds in \citet{Haid18}.
Furthermore, (radiative) stellar feedback can significantly change dust properties and the associated alignment efficiencies (Section~\ref{sec:RAT}). Taken together, these macro- and microphysical changes can significantly alter the findings presented here for pristine MCs, which is why we intend to study the impact of stellar feedback on the observable polarisation in detail in subsequent work. Furthermore, we plan to apply the method presented here to simulations of colliding flows (Joshi et al., submitted; Weis et al., in prep.) and to make an in-depth comparison between the results of both types of simulations.

\section{Conclusions}
\label{sec:conclusion}

We present synthetic dust polarisation maps of 3D-MHD simulations of molecular clouds before the onset of stellar feedback, which form in their larger-scale, galactic environment modelled within the SILCC-Zoom project. The  MHD simulations make use of a chemical network and self-consistently calculate the dust temperature by taking into account radiative shielding.

The dust polarisation radiative transfer (RT) is done with the freely available code POLARIS \citep{Reissl16} and includes a self-consistent calculation of the alignment efficiencies of dust grains with variable sizes. We use radiative torque alignment and present synthetic polarisation observations at wavelengths from 70 $\mu$m to 3 mm at a resolution of 0.12 pc, comparable to recent observations by e.g. the BlastPol experiment at assumed distances of 1 kpc. Our main results are:
\begin{itemize}
 \item Dust grains are well aligned even at high densities (\mbox{$n$ $>$ 10$^3$ cm$^{-3}$}) and visual extinctions ($A_\rmn{V}$ $>$ 1). We demonstrate that the frequently observed depolarisation in molecular clouds is rather caused by strong variations $\left\langle \sigma_{B, \varphi} \right\rangle$ of the magnetic field direction along the LOS.
 \item The observed magnetic field structure, i.e. the polarisation vectors rotated by 90$^\circ$, represents best the mass-weighted, LOS-averaged magnetic field, but is less representative for the volume-weighted, LOS-averaged magnetic field.
 \item The effect of noise can be reduced by convolving the images with a Gaussian beam with a width of few times the pixel size. At $\lambda$ $\gtrsim$ 160 $\mu$m the observed magnetic field traces reliably the underlying field structure in regions with intensities $I$ $\gtrsim$ 2 times the noise level and column densities above \mbox{$\sim$ 1 M$_{\sun}$ pc$^{-2}$}. Typical deviations in these regions are $\lesssim$ 10$^\circ$. 
 In regions of strong column density changes, however, the observed field structure should be considered with caution.
 \item The observed polarisation degree can be related to the variation of the magnetic field along the LOS via an empirical formula
\begin{equation}
 \left\langle \sigma_{B, \varphi} \right\rangle \simeq (55^\circ \pm 12^\circ) - 1.5^\circ \times \frac{p}{1\%} \, .
\end{equation}
For $p$ $<$ 1 -- 2\%, the observed magnetic field orientation is in general not representative for the LOS-averaged magnetic field due to significant variations of the latter around its mean direction. 
 \item Related to that, the differences between the observed and the LOS-averaged magnetic field structure increase with decreasing polarisation degree $p$. For \mbox{$p$ $<$ 2\%}, the typical differences increase to 20$^\circ$ -- 40$^\circ$ and the observed field structure has to be considered with caution.
 \item  The polarisation vectors differ only marginally, by a few 1$^\circ$, for wavelengths $\geq$ 70 $\mu$m allowing for single-wavelength observations to determine the polarisation structure of MCs. Furthermore, the polarisation structure is only little affected by the resolution.
\end{itemize}

We also tested the accuracy of a simplified method to calculate the polarisation structure from simulations, which does not take into account self-consistent alignment efficiencies for dust grains nor RT. It was applied in a number of works  \citep{Lee85,Wardle90,Fiege00,Heitsch01,Padoan01,Kataoka12,Soler13,PlanckXX,King18,Vaisala18}, but has lacked a proof-of-concept so far. By comparing the results of this method with those of the self-consistent dust polarisation RT applied in this work, we find that the method provides a good representation of the observable polarisation structure with typical deviations below 5$^\circ$.

Since this work serves as a proof-of-concept how to combine state-of-the-art MHD simulations, dust alignment theories, and dust polarisation radiative transfer, we limit ourselves to a few selected scientific questions. We plan to extend our work to investigate e.g. the correlation between the magnetic field direction and density structures, the variation of the polarisation degree, or the influence of varying dust properties.

\section*{Acknowledgements}

The authors like to thank the anonymous referee for the comments which helped to significantly improve the paper. 
DS also thanks G. Novak for helpful discussions on the implementation of noise in the polarisation maps.
DS and SW acknowledge the support of the Bonn-Cologne Graduate School, which is funded through the German Excellence Initiative. DS and SW also acknowledge funding by the Deutsche Forschungsgemeinschaft (DFG) via the Collaborative Research Center SFB 956 ``Conditions and Impact of Star Formation'' (subproject C5). SW acknowledges support via the ERC starting grant No. 679852 "RADFEEDBACK". JCIM acknowledges funding via the Priority Program SPP 1573 ``Physics of the Interstellar Medium''.
SR acknowledges support from  the  Deutsche  Forschungsgemeinschaft via the SFB 881 ``The Milky Way System'' (subprojects B1, B2, and B8) and via the Priority Program SPP 1573 ``Physics of the Interstellar Medium'' (grant numbers KL 1358/18.1, KL 1358/19.2).  The FLASH code used in this work was partly developed by the Flash Center for Computational Science at the University of Chicago.
The authors acknowledge the Leibniz-Rechenzentrum Garching for providing computing time on SuperMUC via the project ``pr94du'' as well as the Gauss Centre for Supercomputing e.V. (www.gauss-centre.eu).




\bibliographystyle{mnras}
\bibliography{literature} 



\appendix

\section{Resolution study}
\label{sec:resolution}

\begin{figure*}
 \centering
 \includegraphics[width=\linewidth]{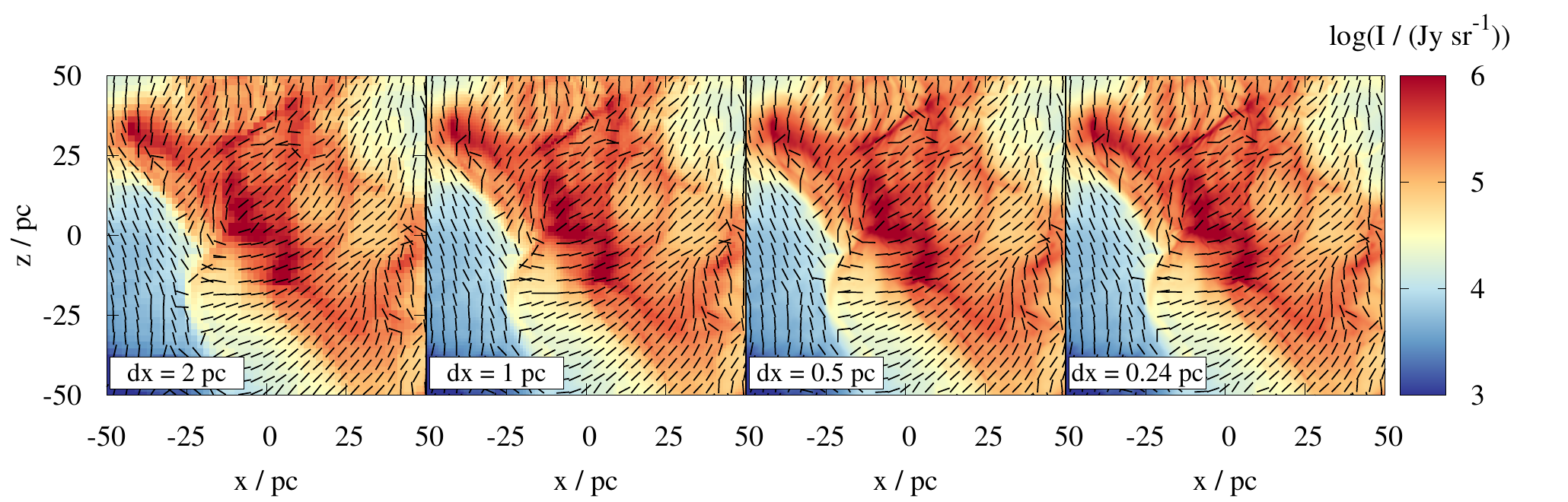}
 \caption{Dust emission intensity and polarisation vectors (black bars) for MC1 at $t_\rmn{evol}$ = 3 Myr along the $z$-direction at $\lambda$ = 1.3 mm for increasing resolution (from left to right). Both, the intensity and the polarisation pattern is remarkably similar for all four resolutions, which indicates that our results shown in the paper are converged.}
 \label{fig:maps_res}
\end{figure*}

In Fig.~\ref{fig:maps_res} we show the maps for the intensity and polarisation vector at $\lambda$ = 1.3 mm for MC1 at $t_\rmn{evol}$ = 3 Myr for different resolutions (see Fig.~\ref{fig:maps} for a resolution of 0.12 pc). We show the case in the absence of noise. We find that the differences in the obtained polarisation pattern are minor, although we vary the resolution by up to a factor of 16.

\begin{figure*}
  \centering
 \includegraphics[width=\textwidth]{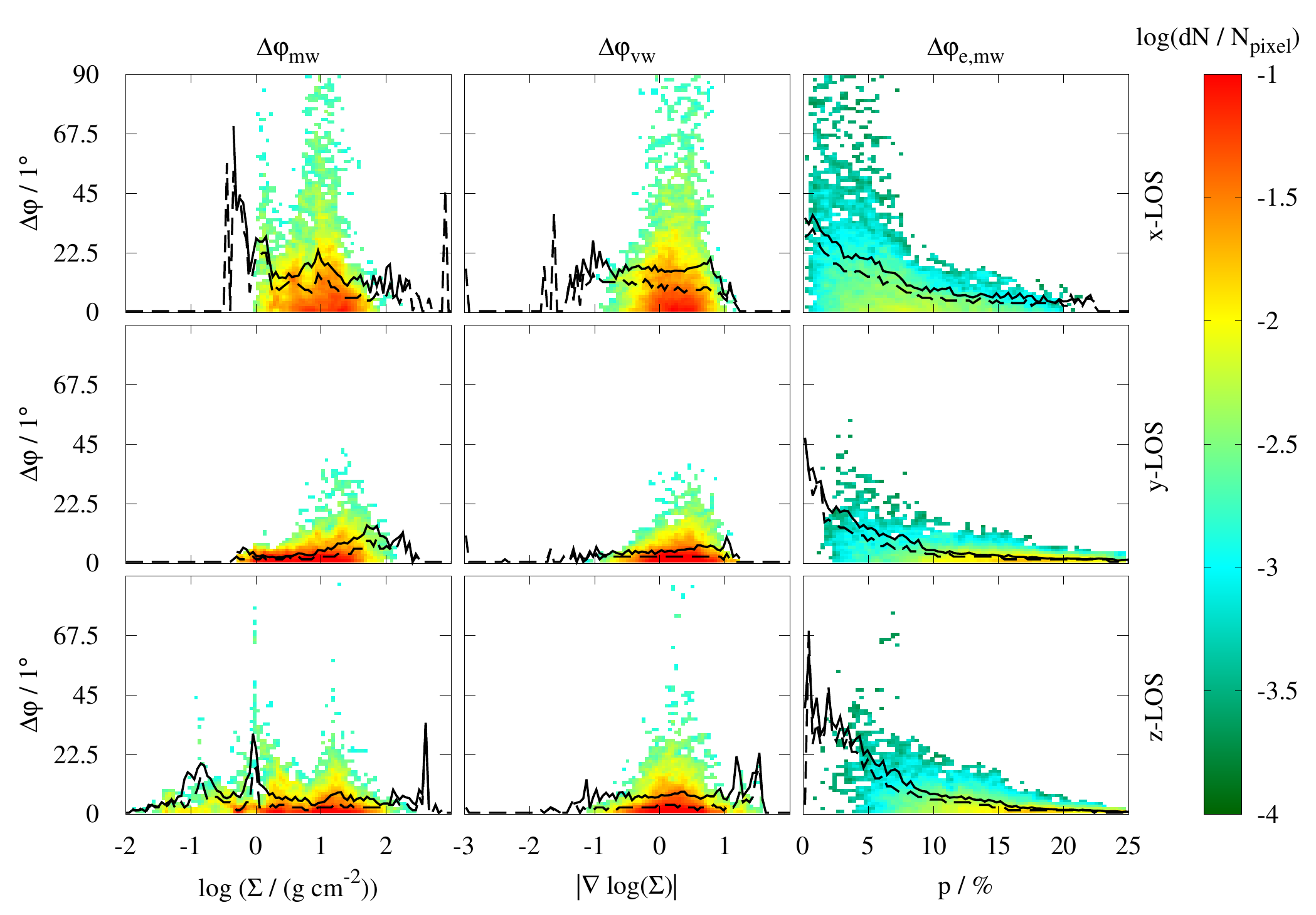}
 \caption{Phase diagram of $\Delta \varphi$ vs $\Sigma$, $|\nabla \Sigma|$, and $p$ (from left to right) but now for a resolution of 1 pc. Despite the expected fact that the scatter in the phase diagram is smaller, the general behaviour as well as the mean (black solid line) and median (black dashed line) values of $\Delta \varphi$ are very similar to that with an 8 times higher resolution (compare the left columns in the Figs.~\ref{fig:dphi_cd}, \ref{fig:dphi_gradcd}, and~\ref{fig:dphi_pol}).}
 \label{fig:dphi_pol_res}
\end{figure*}

Next, we show the phase plots of $\Delta \varphi$ vs $\Sigma$, $|\nabla \Sigma|$, and $p$ for the case without noise in Fig.~\ref{fig:dphi_pol_res} for a resolution of 1 pc. The results are qualitatively and quantitatively similar to the high resolution case (compare to the left column in the Figs.~\ref{fig:dphi_cd}, \ref{fig:dphi_gradcd}, and~\ref{fig:dphi_pol}). As expected, the scatter in the phase diagrams is somewhat lower than for the high-resolution case. However, in particular the mean and median (black lines) are in good agreement with that of the high-resolution run (0.12~pc).

Hence, our results suggest that, when going to lower angular resolution or considering a further distant MC, the obtained polarisation structure is not affected significantly and that the results presented in the paper are converged.

\begin{figure*}
 \centering
 \includegraphics[width=\linewidth]{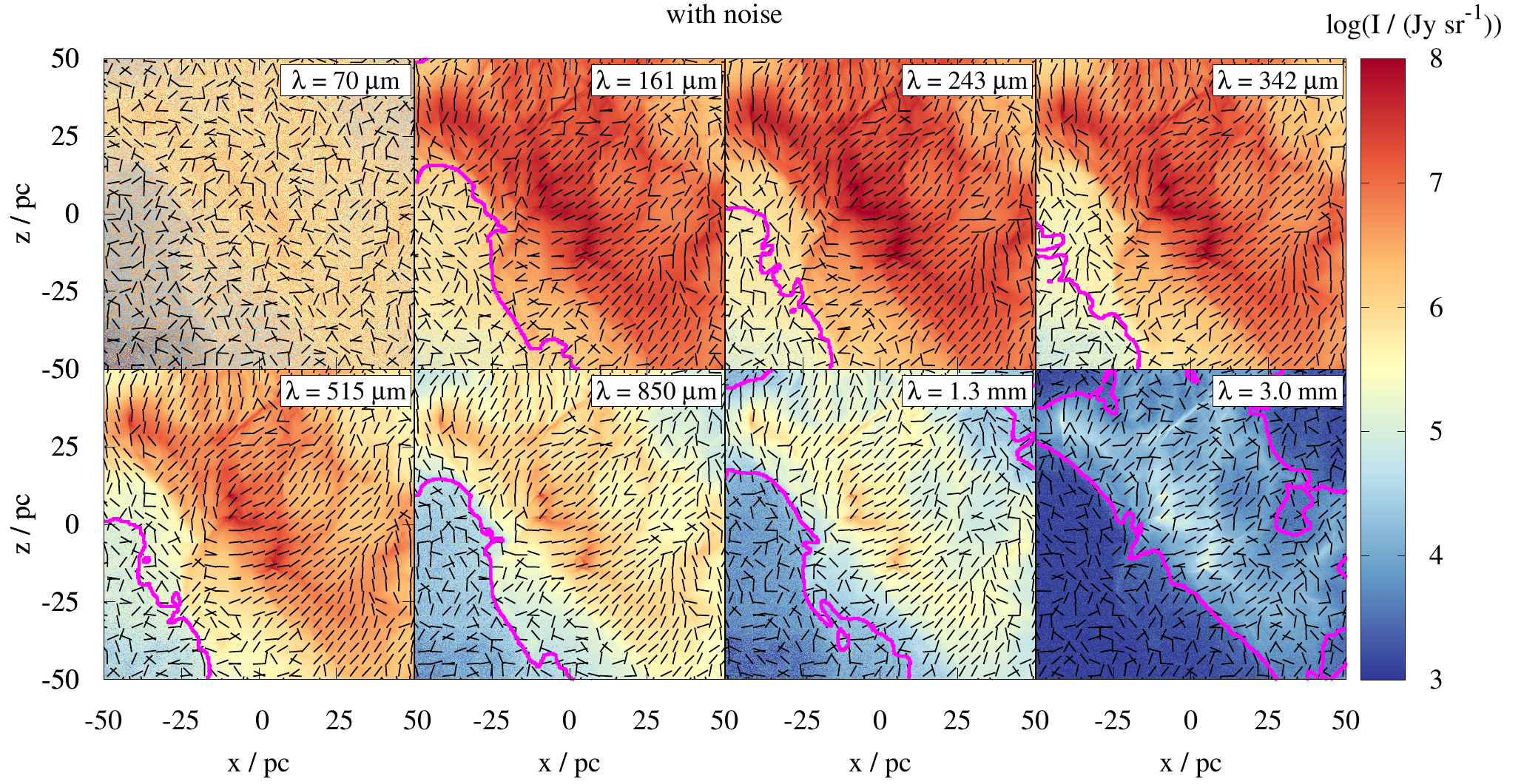}\\
 \includegraphics[width=\linewidth]{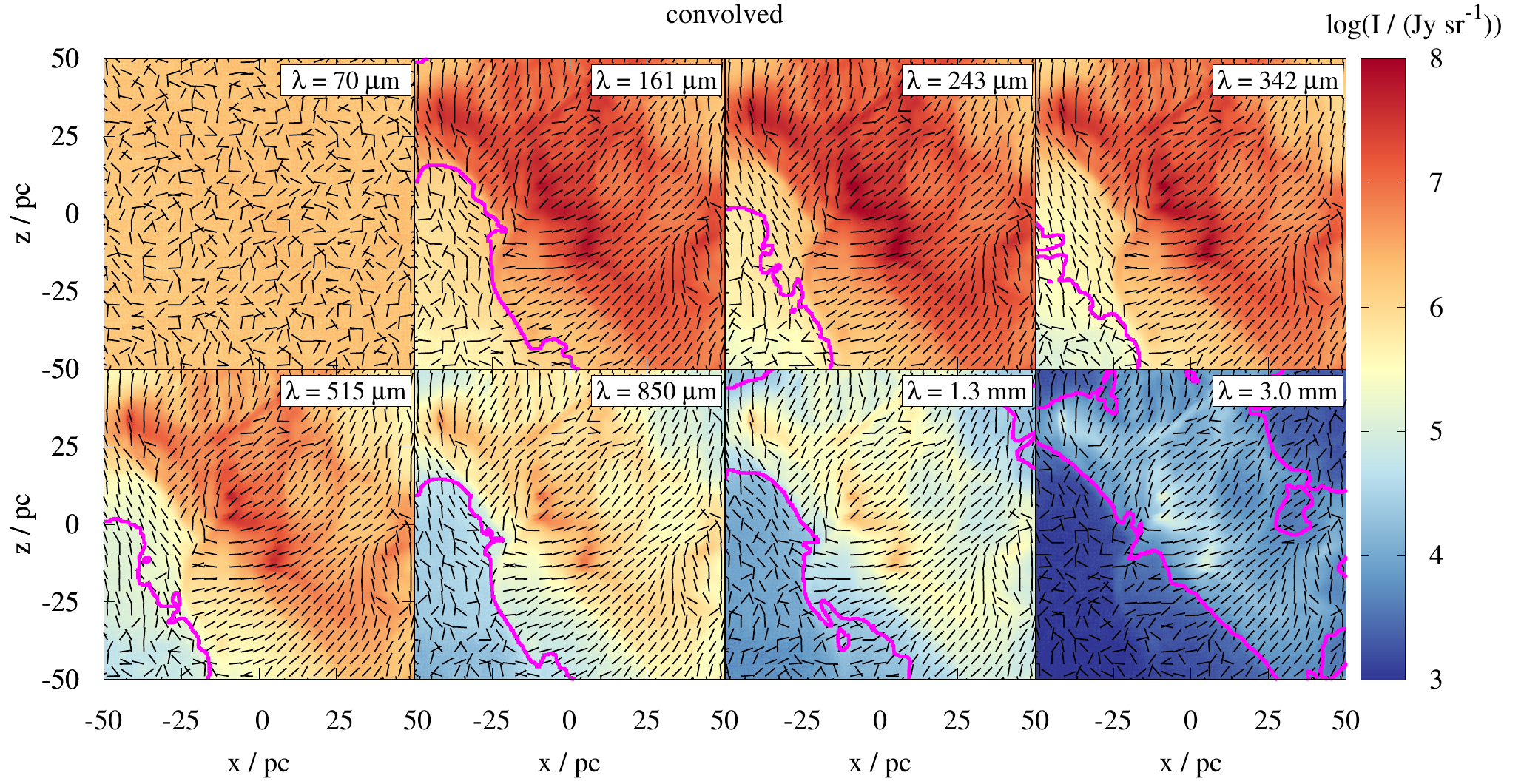}
 \caption{Top: Dust emission intensity and polarisation vectors (black bars) for MC1 at $t_\rmn{evol}$ = 3 Myr along the $z$-direction for all considered wavelengths from 70 $\mu$m to 3 mm (top left to bottom right) including noise calculated according to Eq.~\ref{eq:noise}. The polarisation pattern becomes randomized in particular in the regions of low intensity, whereas in the high-intensity regions the pattern is mostly preserved (compare with Fig.~\ref{fig:maps}). The magenta line shows the 2~$\sigma_I$ contour. Bottom: Same as in the top panel, but now for the convolved image. The convolution improves the accuracy of the polarisation pattern in regions with intensities above a few times the noise level as indicated by the 2~$\sigma_I$ contour (magenta lines). For $\lambda$ = 70 $\mu$m, the noise is dominating and the polarisation pattern remains randomized even after convolution.}
 \label{fig:map_noise_lambda}
\end{figure*}

\section{Wavelength dependent noise}
\label{sec:app_noise}

In Fig.~\ref{fig:map_noise_lambda} (top panel) we show the effect of noise adopted according to Eq.~\ref{eq:noise} and subsequent convolution with a Gaussian beam (bottom panel) for different wavelengths. For all wavelengths expect $\lambda$ = 70~$\mu$m, the observed effect of noise is qualitatively similar to the case of $\lambda$ = 1.3 mm discussed in Section~\ref{sec:noise}. In particular, we find that in regions with \mbox{$I$ $\gtrsim$ 2 $\sigma_I$} (magenta contours) the quality of the polarisation data is significantly improved by the convolution.

For $\lambda$ = 70 $\mu$m, however, the polarisation structure is dominated by noise and cannot be recovered by convolution. This is due to the fact that for $\lambda$ = 70 $\mu$m the overall intensity has dropped in contrast to $\sigma_I$, which increases with decreasing $\lambda$ (Fig.~\ref{fig:noise} and Eq.~\ref{eq:noise}). Also for $\lambda$ = 3~mm, the effect of noise is relatively pronounced, although here the convolution can significantly improve the quality of the polarisation data.

We emphasise that the regions with $I$ $>$ 2 $\sigma_I$ are relatively similar for all $\lambda$ $\gtrsim$ 161 $\mu$m. We attribute this to the fact that in this wavelength range both $I$ and $\sigma_I$ show a similar dependence on $\lambda$ being roughly proportional to $\lambda^{-2}$ as expected in the Rayleigh-Jeans limit.

\bsp	
\label{lastpage}
\end{document}